\documentclass[aps,prd,12pt,preprint,superscriptaddress,nofootinbib,floatfix]{revtex4-1}
\usepackage{latexsym}
\usepackage{amsfonts}
\usepackage[latin1]{inputenc}
\usepackage{subfig}
\usepackage{graphicx}
\usepackage{mathrsfs}
\usepackage{amssymb}
\usepackage{amsmath}
\usepackage{verbatim}
\def\met{{\slash\!\!\!\!\!\:E}_T}

\def\D0{\slash\!\!\!\!\!\!\!\!\!\:D0}

\begin{document}

\begin{flushleft}
{SHEP-10-23}\\
\today
\end{flushleft}

\title{Phenomenology of the minimal $B-L$ extension\\[0.15cm]
 of the Standard Model: the Higgs sector}
\vspace*{1.0truecm}
\author{Lorenzo Basso}
\affiliation{
School of Physics \& Astronomy, University of Southampton,\\
Highfield, Southampton SO17 1BJ, UK}
\affiliation{
Particle Physics Department, Rutherford Appleton Laboratory, \\Chilton,
Didcot, Oxon OX11 0QX, UK}
\author{Stefano Moretti}
\affiliation{
School of Physics \& Astronomy, University of Southampton,\\
Highfield, Southampton SO17 1BJ, UK}
\affiliation{
Particle Physics Department, Rutherford Appleton Laboratory, \\Chilton,
Didcot, Oxon OX11 0QX, UK}
\author{Giovanni Marco Pruna}
\affiliation{
School of Physics \& Astronomy, University of Southampton,\\
Highfield, Southampton SO17 1BJ, UK}
\affiliation{
Particle Physics Department, Rutherford Appleton Laboratory, \\Chilton,
Didcot, Oxon OX11 0QX, UK}
\begin{abstract}
{\small \noindent
We investigate the phenomenology of the Higgs sector of the minimal $B-L$ extension of the 
Standard Model. We present results for both the foreseen energy stages of the Large Hadron Collider 
($\sqrt s=7$ and 14 TeV). We show that in such a scenario several novel production and decay channels
involving the two physical Higgs states could be accessed at such a machine. Amongst these,
several Higgs signatures have very distinctive features
with respect to those of other models with enlarged Higgs sector, as they involve interactions of 
Higgs bosons between themselves, with $Z'$ bosons as well as with heavy neutrinos. }
\end{abstract}
\maketitle

\newpage


\section{Introduction}
\label{Sec:Intro}
In the past years, major efforts has been devoted to the realisation
of the Large Hadron Collider (LHC), the largest and
most powerful running collider in the world. One of its scopes
is discovering the means of generating
masses for all known (and possibly new) particles.

As a matter of fact, while it is widely accepted that the way of
realising the aforementioned mass generation is represented by the
Higgs Mechanism, there is still no experimental evidence of any Higgs
boson.

As for the models implementing the Higgs mechanism, the Standard Model ($SM$) 
is based on just one complex Higgs
doublet consisting of four degrees of freedom, three of which, after
spontaneous Electro-Weak Symmetry Breaking ($EWSB$), turn out to be
absorbed in the longitudinal polarisation component of each of the
three weak gauge bosons, $W^{\pm}$ and $Z$, whilst the fourth one
gives the physical Higgs state $h$ (for a detailed ``anatomy'' of the
Higgs mechanism in the $SM$ see \cite{Djouadi:2005gi}).

Despite the $SM$ provides a beautiful explanation for most known particle
phenomena, it turns out to be unsatisfactory from several points of
view. Apart from some feeble hints of the $SM$ inadequacy coming from
precision tests, it does not produce a viable dark matter candidate, it
does not incorporate dark energy, it does not provide enough CP violation 
to explain the baryonic matter-antimatter asymmetry of the universe and, finally,
it cannot describe the experimentally observed
evidence of neutrino oscillations.

To stay with the latter aspect, and following a bottom-up approach,
one can attempt to remedy this issue
through a minimal extension of the $SM$: the so-called minimal $B-L$
model (see \cite{Jenkins:1987ue,Buchmuller:1991ce} and
        \cite{Khalil:2006yi}). Such a scenario consists of
a further $U(1)_{B-L}$ gauge group in addition the $SM$ gauge structure, 
three right-handed neutrinos (designed to cancel anomalies) and
an additional complex Higgs singlet responsible for giving mass to an
additional $Z'$ gauge boson. Therefore, the scalar sector is made of
two real CP-even scalars, that will mix together.

In this theoretical framework, following the $B-L$ symmetry breaking,
the right-handed neutrinos can acquire a Majorana mass of the order of
the TeV scale ($\sim$ $B-L$ symmetry breaking Vacuum Expectation Value
(VEV)), and this can in turn explain the smallness of the light-neutrinos
masses via the Type I see-saw mechanism
(see \cite{Minkowski:1977sc,VanNieuwenhuizen:1979hm,Yanagida:1979as,GellMann:1980vs,S.L.Glashow,Mohapatra:1979ia}).


Finally, it is important to note that in this model the ${B-L}$ breaking can
take place at the TeV scale, i.e., far below that of any Grand Unified
Theory (GUT), thereby giving rise to new and interesting phenomenology
at present and future particle accelerators
\cite{Emam:2007dy,Basso:2008iv,Emam:2008zz,Huitu:2008gf,Basso:2009hf,Basso:2010pe}.

In the present work we study the phenomenology at the Large Hadron
Collider (LHC) of the scalar sector of the minimal $B-L$ model.
We will present production cross sections, Branching Ratios (BRs) and
event rates for the $B-L$ Higgs bosons, highlighting the analogies and
differences
with respect to the $SM$ case and other models that show a similar
phenomenology in the Higgs sector (as the scalar singlet extension of
the $SM$,
see \cite{BahatTreidel:2006kx,O'Connell:2006wi,Barger:2007im,Bhattacharyya:2007pb,Profumo:2007wc,Barger:2008jx}),
and
we will use these results to introduce new Higgs boson signatures at
the LHC, that could be the hallmark of the model considered here:
e.g., four lepton decays of a heavy Higgs boson via pairs of $Z'$
gauge bosons (which, e.g., in the $SM$ also occurs via $W^+W^-$ and $ZZ$
but in very different kinematic regions), light Higgs boson pair
production via the heavy Higgs boson (forbidden, e.g., over the
currently allowed parameter space of the Minimal Supersymmetric
Standard Model ($MSSM$)) and heavy neutrino pair production via a light
Higgs boson (yielding, e.g., very exotic and clean like-sign dilepton
signatures, with or without jets).

This work can be seen as the continuation of the studies started in
Refs.~\cite{Basso:2008iv,Basso:2009hf,Basso:2010pe}, 
where we dealt with the other new sectors of the model (i.e., the $Z'$
gauge boson and the heavy neutrino ones), 
and relies on the results of
Refs.~\cite{Basso:2010jt,Basso:2010jm,Basso:2010hk} where
the Higgs parameter space of the minimal $B-L$ model was studied in
detail by accounting for all experimental and theoretical
constraints.

This paper is organised as follows: in the next section we describe
the model in its relevant (to this study) parts, in the following
one we describe the details of the analysis carried out, in
section \ref{Sec:Results} we present our numerical results, then we
conclude in section \ref{Sec:Conclusions}.


\section{The model}
\label{Sec:Model}
The model under study is the minimal $U(1)_{B-L}$ extension of the $SM$
(see Refs.~\cite{Basso:2008iv,Basso:2010jt,Basso:2010jm} for
conventions and references), in which the $SM$ gauge group is augmented
by a $U(1)_{B-L}$ factor, related to the Baryon minus Lepton ($B-L$)
gauged number. In the complete model, the classical gauge invariant
Lagrangian, obeying the $SU(3)_C\times SU(2)_L\times U(1)_Y\times
U(1)_{B-L}$ gauge symmetry, can be decomposed as:
\begin{equation}\label{L}
\mathscr{L}=\mathscr{L}_s + \mathscr{L}_{YM} + \mathscr{L}_f + \mathscr{L}_Y \, .
\end{equation}
\noindent
The scalar Lagrangian is:
\begin{equation}\label{new-scalar_L}
\mathscr{L}_s=\left( D^{\mu} H\right) ^{\dagger} D_{\mu}H + 
\left( D^{\mu} \chi\right) ^{\dagger} D_{\mu}\chi - V(H,\chi ) \, ,
\end{equation}
with the scalar potential given by
\begin{eqnarray}\nonumber
V(H,\chi )&=&m^2H^{\dagger}H + \mu ^2\mid\chi\mid ^2
					+ \left( \begin{array}{cc} 
					H^{\dagger}H& \mid\chi\mid
					^2\end{array}\right) 
					\left( \begin{array}{cc} \lambda
					_1 & \frac{\lambda _3}{2} \\ 
			 \frac{\lambda _3}{2} & \lambda _2 \\ \end{array} \right) \left( \begin{array}{c} H^{\dagger}H \\ \mid\chi\mid ^2 \\ \end{array} \right)\\
			  \nonumber \\ \label{BL-potential}
		&=&m^2H^{\dagger}H + \mu ^2\mid\chi\mid ^2 + \lambda _1 (H^{\dagger}H)^2 +\lambda _2 \mid\chi\mid ^4 + \lambda _3 H^{\dagger}H\mid\chi\mid ^2  \, ,
\end{eqnarray}
where $H$ and $\chi$ are the complex scalar Higgs doublet and singlet fields, respectively.

We generalise the $SM$ discussion of spontaneous $EWSB$ to the more complicated classical potential of eq.
(\ref{BL-potential}). To determine the condition for $V(H,\chi )$ to be bounded from below, it is sufficient to study its behaviour
for large field values, controlled by the matrix in the first line of eq. (\ref{BL-potential}). Requiring such a matrix to be 
positive-definite, we obtain the conditions:
\begin{equation}\label{inf_limitated}
4 \lambda _1 \lambda _2 - \lambda _3^2>0 \, ,
\end{equation}
\begin{equation}\label{positivity}
\lambda _1, \lambda _2 > 0 \, .
\end{equation}
If the above conditions are satisfied, we can proceed to the minimisation of $V$ as a function of constant 
VEVs for the two Higgs
fields. Making use of gauge invariance, it is not restrictive to assume:
\begin{equation}\label{min}
\left< H \right> \equiv \left( \begin{array}{c} 0 \\ \frac{v}{\sqrt{2}} \end{array} \right)\, , 
	\hspace{2cm} \left< \chi \right> \equiv \frac{x}{\sqrt{2}}\, ,
\end{equation} 
with $v$ and $x$ real and non-negative. The physically most interesting solutions to the minimisation of eq.~(\ref{BL-potential}) are obtained for $v$ and $x$ both non-vanishing:
\begin{eqnarray}\label{sol_min1}
v^2 &=& \frac{-\lambda _2 m^2 + \frac{\lambda _3}{2}\mu ^2}{\lambda _1 \lambda _2 - \frac{\lambda _3^{\phantom{o}2}}{4}} \, ,\\
\nonumber  \\ \label{sol_min2}
x^2 &=& \frac{-\lambda _1 \mu ^2 + \frac{\lambda _3}{2}m ^2}{\lambda _1 \lambda _2 - \frac{\lambda _3^{\phantom{o}2}}{4}} \, .
\end{eqnarray}

To compute the scalar masses, we must expand the potential in eq. (\ref{BL-potential}) around the minima
in eqs. (\ref{sol_min1}) and (\ref{sol_min2}).
We denote by
$h_1$ and $h_2$ the scalar fields of definite masses, $m_{h_1}$ and $m_{h_2}$ respectively, and we conventionally choose
$m^2_{h_1} < m^2_{h_2}$. After standard manipulations, the explicit expressions for the scalar mass eigenvalues and eigenvectors are:
\begin{eqnarray}\label{mh1}
m^2_{h_1} &=& \lambda _1 v^2 + \lambda _2 x^2 - \sqrt{(\lambda _1 v^2 - \lambda _2 x^2)^2 + (\lambda _3 xv)^2} \, ,\\ \label{mh2}
m^2_{h_2} &=& \lambda _1 v^2 + \lambda _2 x^2 + \sqrt{(\lambda _1 v^2 - \lambda _2 x^2)^2 + (\lambda _3 xv)^2} \, ,
\end{eqnarray}
\begin{equation}\label{scalari_autostati_massa}
\left( \begin{array}{c} h_1\\h_2\end{array}\right) = \left( \begin{array}{cc} \cos{\alpha}&-\sin{\alpha}\\ \sin{\alpha}&\cos{\alpha}
	\end{array}\right) \left( \begin{array}{c} h\\h'\end{array}\right) \, ,
\end{equation}
where $-\frac{\pi}{2}\leq \alpha \leq \frac{\pi}{2}$ fulfils \footnote{In all generality, the whole interval $0\leq \alpha < 2\pi$ is halved
because an orthogonal transformation is invariant under $\alpha \rightarrow \alpha + \pi$. We could re-halve the interval by noting
that it is invariant also under $\alpha \rightarrow -\alpha$ if we permit the eigenvalues inversion, but this is forbidden by our
convention $m^2_{h_1} < m^2_{h_2}$. Thus $\alpha$ and $-\alpha$ are independent solutions.}:\label{scalar_angle}
\begin{eqnarray}\label{sin2a}
\sin{2\alpha} &=& \frac{\lambda _3 xv}{\sqrt{(\lambda _1 v^2 - \lambda _2 x^2)^2 + (\lambda _3 xv)^2}} \, ,\\ \label{cos2a}
\cos{2\alpha} &=& \frac{\lambda _1 v^2 - \lambda _2 x^2}{\sqrt{(\lambda _1 v^2 - \lambda _2 x^2)^2 + (\lambda _3 xv)^2}}\, .
\end{eqnarray}

For our numerical study of the extended Higgs sector, it is useful to invert eqs.~(\ref{mh1}), (\ref{mh2}) and (\ref{sin2a}), 
to extract the parameters in the Lagrangian in terms of the physical quantities $m_{h_1}$, $m_{h_2}$ and $\sin{2\alpha}$:
\begin{eqnarray}\nonumber
\lambda _1 &=& \frac{m_{h_2}^2}{4v^2}(1-\cos{2\alpha}) + \frac{m_{h_1}^2}{4v^2}(1+\cos{2\alpha}),\\ \nonumber
\lambda _2 &=& \frac{m_{h_1}^2}{4x^2}(1-\cos{2\alpha}) + \frac{m_{h_2}^2}{4x^2}(1+\cos{2\alpha}),\\ \label{inversion}
\lambda _3 &=& \sin{2\alpha} \left( \frac{m_{h_2}^2-m_{h_1}^2}{2xv} \right).
\end{eqnarray}

Moving to the Yang-Mills Lagrangian $\mathscr{L}_{YM}$, the non-Abelian field strengths therein are the same as in the $SM$ whereas the Abelian ones can be written as follows:
\begin{equation}\label{La}
\mathscr{L}^{\rm Abel}_{YM} = 
-\frac{1}{4}F^{\mu\nu}F_{\mu\nu}-\frac{1}{4}F^{\prime\mu\nu}F^\prime _{\mu\nu}\, ,
\end{equation}
where
\begin{eqnarray}\label{new-fs3}
F_{\mu\nu}		&=&	\partial _{\mu}B_{\nu} - \partial _{\nu}B_{\mu} \, , \\ \label{new-fs4}
F^\prime_{\mu\nu}	&=&	\partial _{\mu}B^\prime_{\nu} - \partial _{\nu}B^\prime_{\mu} \, .
\end{eqnarray}
In this field basis, the covariant derivative is:
\begin{equation}\label{cov_der}
D_{\mu}\equiv \partial _{\mu} + ig_S T^{\alpha}G_{\mu}^{\phantom{o}\alpha} 
+ igT^aW_{\mu}^{\phantom{o}a} +ig_1YB_{\mu} +i(\widetilde{g}Y + g_1'Y_{B-L})B'_{\mu}\, .
\end{equation}
The ``pure'' or ``minimal'' $B-L$ model is defined by the condition $\widetilde{g} = 0$, that implies no mixing between the $B-L$ $Z'$ and $SM$ $Z$ gauge bosons.

The fermionic Lagrangian (where $k$ is the
generation index) is given by
\begin{eqnarray} \nonumber
\mathscr{L}_f &=& \sum _{k=1}^3 \Big( i\overline {q_{kL}} \gamma _{\mu}D^{\mu} q_{kL} + i\overline {u_{kR}}
			\gamma _{\mu}D^{\mu} u_{kR} +i\overline {d_{kR}} \gamma _{\mu}D^{\mu} d_{kR} +\\
			  && + i\overline {l_{kL}} \gamma _{\mu}D^{\mu} l_{kL} + i\overline {e_{kR}}
			\gamma _{\mu}D^{\mu} e_{kR} +i\overline {\nu _{kR}} \gamma _{\mu}D^{\mu} \nu
			_{kR} \Big)  \, ,
\end{eqnarray}
 where the fields' charges are the usual $SM$ and $B-L$ ones (in
  particular, $B-L = 1/3$ for quarks and $-1$ for leptons {with no
  distinction between generations, hence ensuring universality)}. The
  $B-L$ charge assignments of the fields as well as the introduction
  of new fermionic right-handed heavy neutrinos ($\nu_R$'s) and a
  scalar Higgs field ($\chi$, with charge $+2$ under $B-L$) are
  generally designed to  
  ensure the gauge invariance of the theory. Moreover, as we have
  already mentioned
  in section \ref{Sec:Intro}, the heavy neutrinos have also the aim of
  eliminating the triangular $B-L$ gauge anomalies.
  Therefore, a $B-L$  gauge extension of the $SM$ gauge group
  broken at the TeV scale requires
  at least one new scalar field and three new fermionic fields which are
  charged with respect to the $B-L$ group.

Finally, the Yukawa interactions are:
\begin{eqnarray}\nonumber
\mathscr{L}_Y &=& -y^d_{jk}\overline {q_{jL}} d_{kR}H 
                 -y^u_{jk}\overline {q_{jL}} u_{kR}\widetilde H 
		 -y^e_{jk}\overline {l_{jL}} e_{kR}H \\ \label{L_Yukawa}
	      & & -y^{\nu}_{jk}\overline {l_{jL}} \nu _{kR}\widetilde H 
	         -y^M_{jk}\overline {(\nu _R)^c_j} \nu _{kR}\chi +  {\rm 
h.c.}  \, ,
\end{eqnarray}
{where $\widetilde H=i\sigma^2 H^*$ and  $i,j,k$ take the values $1$ to $3$},
where the last term is the Majorana contribution and the others the usual Dirac ones.

Neutrino mass eigenstates, obtained after applying the see-saw
mechanism, will be called $\nu_l$ (with $l$ standing for light) and
$\nu_h$ (with $h$ standing for heavy), where the first ones are the
$SM$-like ones. With a reasonable choice of Yukawa couplings, the heavy
neutrinos can have masses $m_{\nu_h} \sim \mathcal{O}(100)$ GeV.


\section{Analysis details}
\label{Sec:Analysis_details}
As spelled out already, the independent physical parameters of the
Higgs sector of the scenario considered here are 
\begin{itemize}
\item $m_{h_1}$, $m_{h_2}$ and $\alpha$, the Higgs boson masses and
  mixing angle.
We will span over continuous intervals in the case of the first two
quantities while 
adopting discrete values for the third one. Masses and couplings
(which depend on the Higgs 
mixing) have been tested against the experimental limits
obtained at the Large Electron-Positron (LEP) collider and at the
Tevatron.
\end{itemize}

\noindent
In order to explore efficiently the expanse of parameter space
pertaining to the minimal $B-L$ model,
we introduce two extreme conditions, which makes the model intuitive,
though at the end it should be 
borne in mind that intermediate solutions are most probable.
The two conditions are obtained by setting:
\begin{enumerate}
\item $\alpha=0$, this is the decoupling limit, with $h_1$ behaving
  like the $SM$ Higgs.  
\item $\alpha=\frac{\pi}{2}$, which is the so-called inversion limit,
  in which $h_2$ is the $SM$ 
Higgs  (though recall that this possibility is phenomenologically not
viable, see \cite{Dawson:2009yx} for a complete analysis in the Higgs
singlet extension context).  
\end{enumerate}

Furthermore, concerning the strength of Higgs interactions, some of the salient 
phenomenological behaviours can be summarised as follows:

\vspace*{0.1cm}
{--} $SM$-like interactions scale with $\cos{\alpha}$($\sin{\alpha}$)
for $h_1$($h_2$);

\vspace*{0.1cm}
{--} those involving the other new $B-L$ fields, like $Z'$ and heavy
neutrinos, scale with  
the complementary angle, i.e., with $\sin{\alpha}$($\cos{\alpha}$) for
$h_1$($h_2$); 

\vspace*{0.1cm}
{--} triple (and quadruple) Higgs couplings are possible and can
induce resonant behaviours, so that, e.g.,  
the $h_2\rightarrow h_1\,h_1$ decay can become dominant if $m_{h_2}>2m_{h_1}$.
\vspace*{0.25cm}

Other than $m_{h_1}$, $m_{h_2}$ and $\alpha$, additional parameters
are the following.  
\begin{itemize}
\item 
$g'_1$, the new $U(1)_{B-L}$ gauge  coupling. We will adopt discrete
  perturbative values for this quantity. 
\item $M_{Z'}$, the new gauge boson mass.
An indirect constraint on $M_{Z'}$ comes from analyses at LEP of
precision $EW$ data (see \cite{Cacciapaglia:2006pk}, based on
the analysis of experimental data published in
\cite{Anthony:2003ub,ew:2003ih,Azzi:2004rc,Woods:2004zr,Z-Pole})
\footnote{A less
  conservative approach, based on Fermi-type effective four-fermions
  interactions, gives the weaker constraint $\frac{M_{Z'}}{g'_1} \geq
  6\; \rm{TeV}$ \cite{Carena:2004xs}.}:

\begin{equation}\label{LEP_bound}
\frac{M_{Z'}}{g'_1} \geq 7\; \rm{TeV}\, .
\end{equation}
Further limits have been obtained at Tevatron
\cite{Aaltonen:2008vx,Aaltonen:2008ah,Basso:2010pe}. Both have been
taken into account here.
\item$m_{\nu _h}$, the heavy neutrino masses. We take them to be 
degenerate and relatively
       light.
\item $m_{\nu _l}$, the $SM$ (or light) neutrino masses.  We use the
cosmological upper bound $\sum_l m_{\nu _l}<1$ eV
\cite{Fogli:2008ig}. Ultimately, they have been taken to be
$m_{\nu_l}=10^{-2}$ eV. 

(For illustrative purposes we take all neutrino masses, both light and
heavy, to be degenerate.)

\end{itemize}

Notice that the theoretical limits from vacuum stability, triviality
and perturbative unitarity obtained in
Refs.~\cite{Basso:2010jt,Basso:2010jm,Basso:2010hk} were all taken
into account here.

In this paper we will consider only the qualitative results of the
analysis of the $EW$ precision constraints made in
\cite{Dawson:2009yx} in the context of singlet scalar extensions of
the $SM$ (we assume that the inversion limit is not phenomenologically
allowed), though we would like to mention here the fact that in our
model, due to the different particle content, the constraints on the
precision parameters can be significantly altered (because of, e.g.,
the presence of heavy neutrinos and the $Z'$ gauge boson in the
definition of the $EW$ precision parameters). In the following, we
will not investigate these aspects any further.

The numerical analysis was performed with CalcHEP~\cite{Pukhov:2004ca} 
with the model introduced through LanHEP~\cite{Semenov:1996es}. This
implementation was described at length in Ref.~\cite{Basso:2008iv}, so
we refer the reader to that publication. A version of the model 
somewhat improved with respect to the one
discussed in Ref.~\cite{Basso:2008iv} has been used for this work
though. Here are the differences.

\begin{itemize}
\item The one-loop vertices $g-g-h_1(h_2)$, $\gamma -\gamma-h_1(h_2)$ and $\gamma -Z(Z')-h_1(h_2)$ via $W$ gauge bosons and heavy quarks (top, bottom and charm) have been implemented, adapting the formulae in Ref~\cite{Gunion:1989we}.
\item Running masses for top, bottom and charm quarks, evaluated at the Higgs boson mass: $Q= m_{h_1}(m_{h_2})$, depending on which scalar boson is involved in the interaction.
\item Running of the QCD coupling constant, at two-loops with $5$ active flavours.
\end{itemize}
Finally, the NLO QCD $k$-factor for the gluon fusion process \cite{Graudenz:1992pv,Spira:1995rr,Djouadi:2005gi}\footnote{Notice 
that in Ref.~\cite{Spira:1995rr} (Ref.~\cite{Djouadi:2005gi}), $m_t=174$($178$) GeV, while we used $m_t=172.5$ GeV as top-quark pole mass value.} has been used. Regarding the other processes, 
we decided to not implement their $k$-factors since they are much smaller in comparison.


\section{Results}
\label{Sec:Results}
In this section we present our results for the scalar sector of the $B-L$ model. We first present 
cross-sections at $\sqrt{s}=7$ and $14$ TeV for the two Higgs bosons, as well as their BRs, 
for some fixed values of the scalar mixing angle $\alpha$. Its values have been chosen in each plot to 
highlight some relevant phenomenological aspects. We will then focus on some phenomenologically viable signatures.

\subsection{Standard production mechanisms}
\begin{figure}[!h]
  \subfloat[]{ 
  \label{xs_h1_7}
  \includegraphics[angle=0,width=0.48\textwidth ]{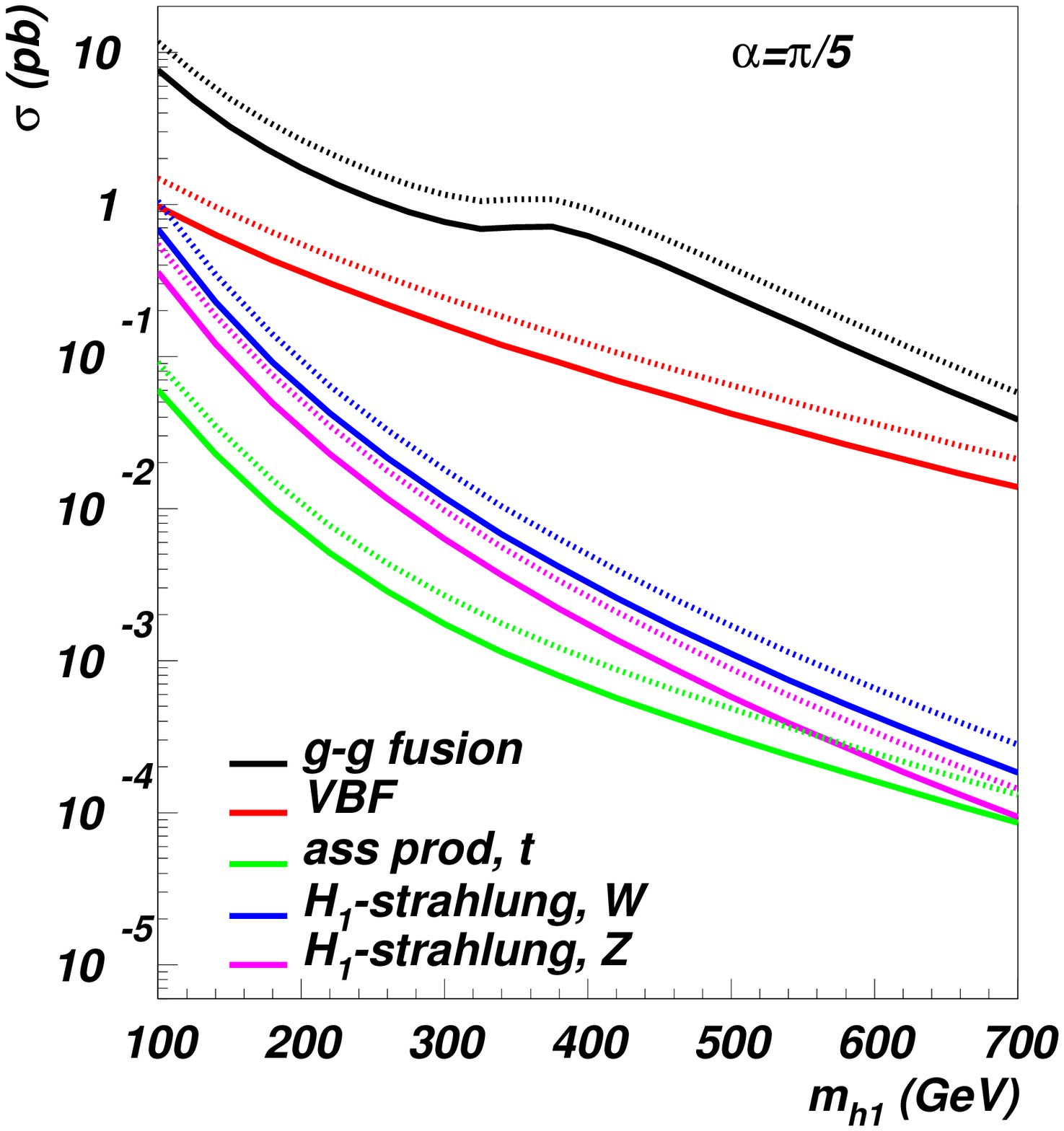}}
  \subfloat[]{
  \label{xs_h2_7}
  \includegraphics[angle=0,width=0.48\textwidth ]{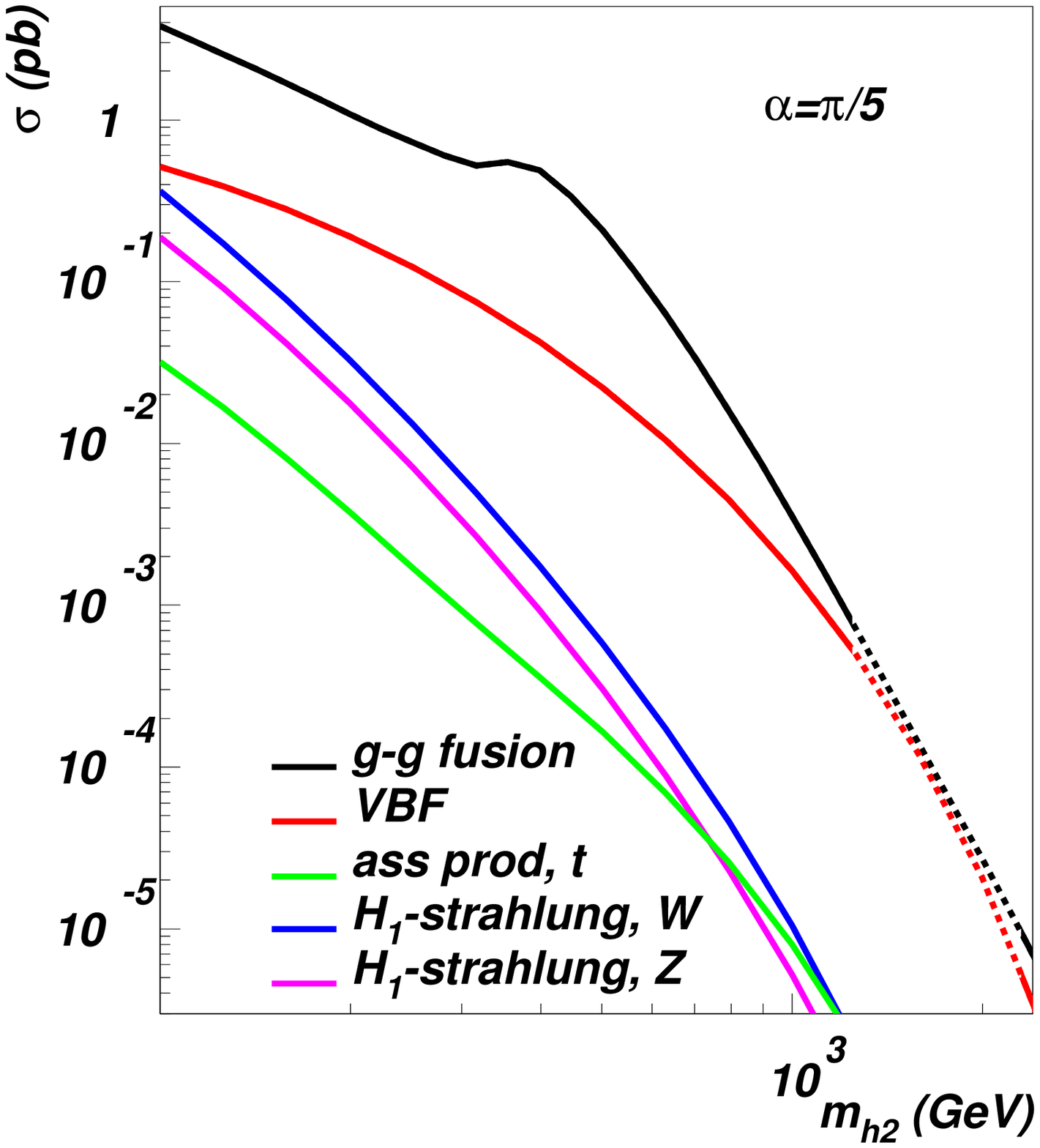}}\\
  \subfloat[]{
  \label{xs_h1_14}
  \includegraphics[angle=0,width=0.48\textwidth ]{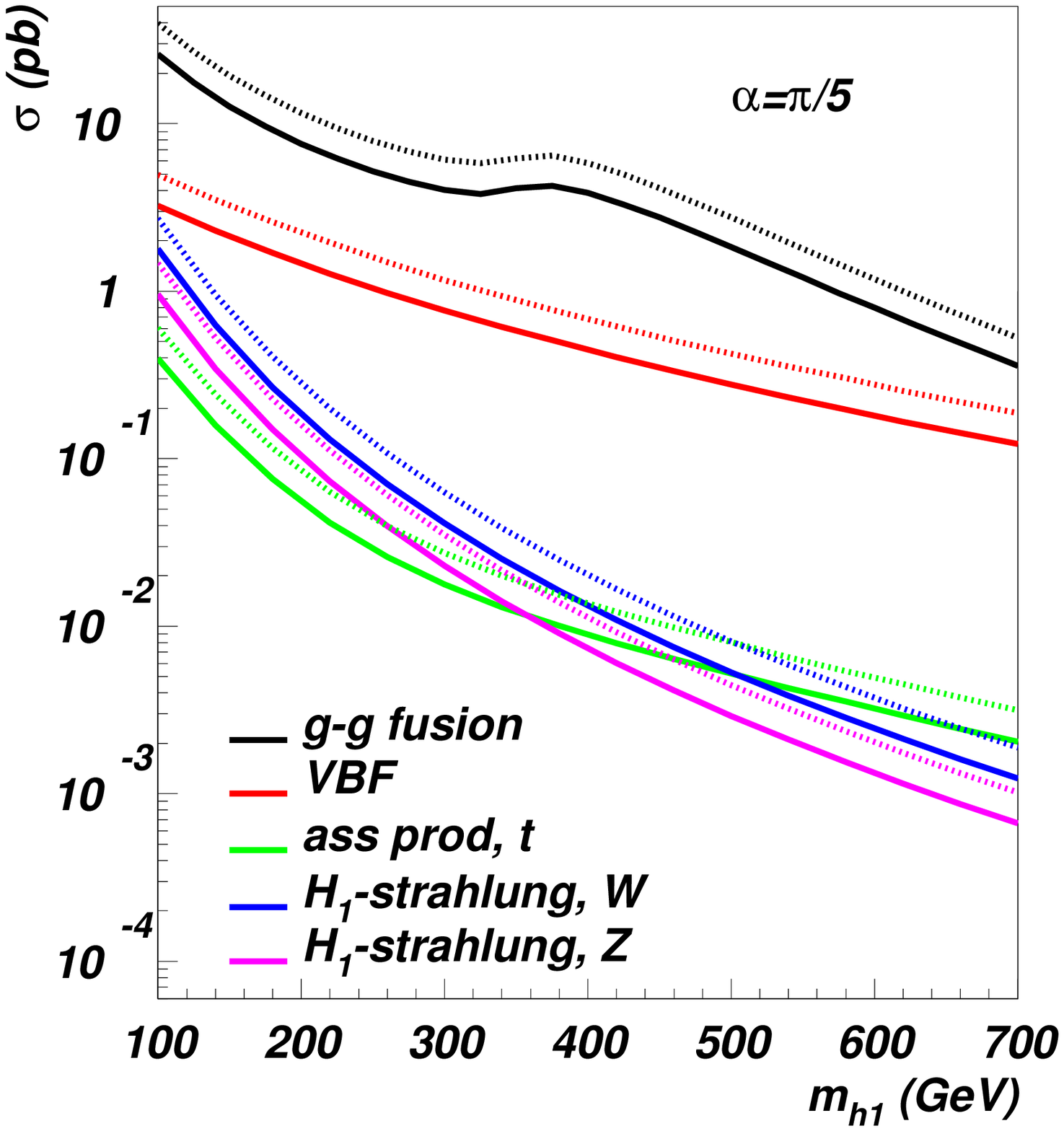}}  
  \subfloat[]{
  \label{xs_h2_14}
  \includegraphics[angle=0,width=0.48\textwidth ]{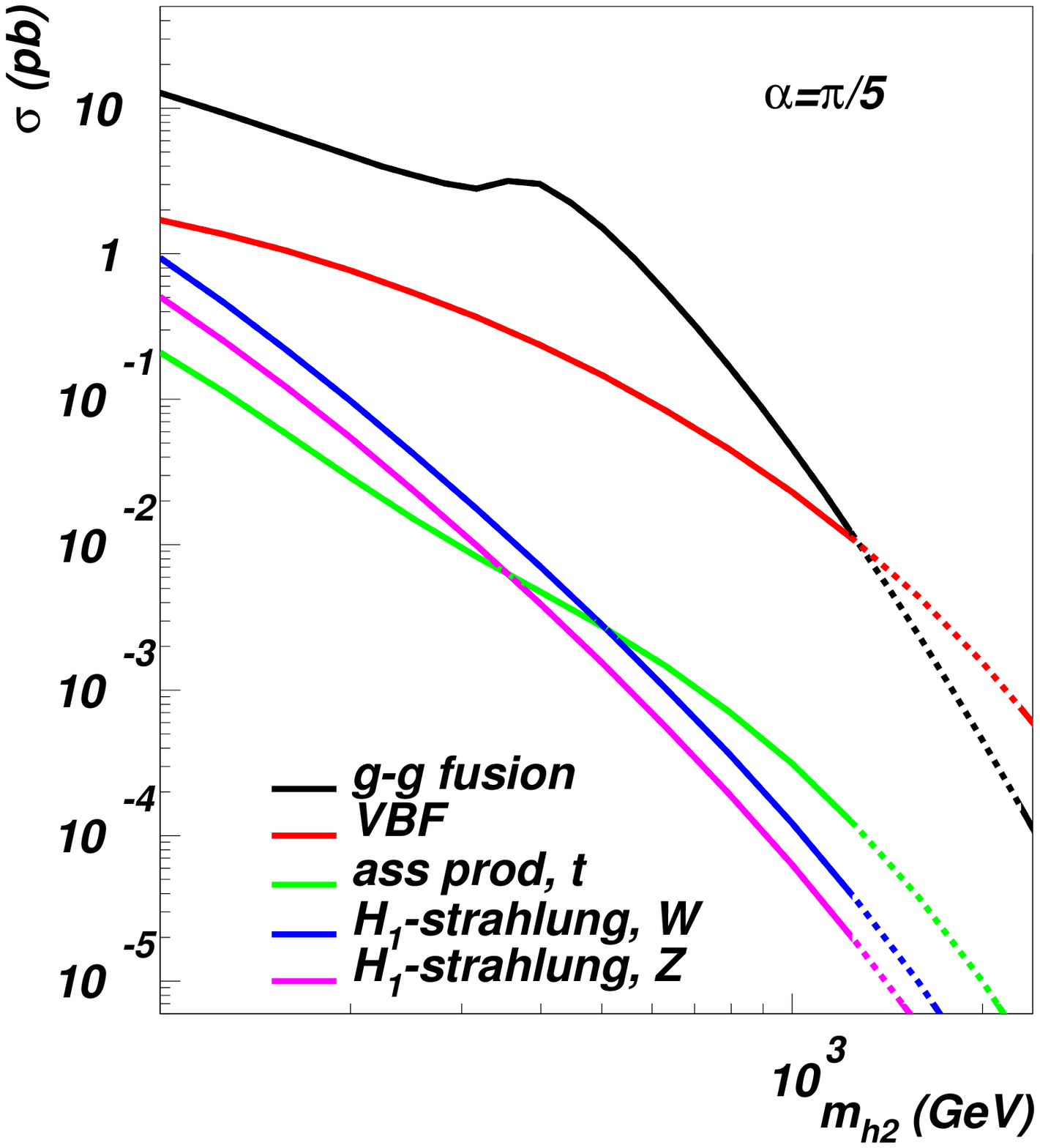}}  
  \vspace*{-0.5cm}
  \caption{\it Cross-sections in the $B-L$ model for $h_1$ at the LHC
    (\ref{xs_h1_7}) at $\sqrt{s}=7$ TeV and (\ref{xs_h1_14}) at
    $\sqrt{s}=14$ TeV, and for $h_2$ (\ref{xs_h2_7}) at $\sqrt{s}=7$
    TeV and (\ref{xs_h2_14}) at $\sqrt{s}=14$ TeV. Dashed lines in
    figs.~(\ref{xs_h1_7}) and (\ref{xs_h1_14}) refer to $\alpha
    =0$. The dotted part of the lines in fig.~(\ref{xs_h2_14}) refer
    to $h_2$ masses excluded by Unitarity (see
    Ref.~\cite{Basso:2010jt}).}
  \label{Xs}
\end{figure}

In figure~\ref{Xs} we present the cross-sections for the most relevant
production mechanisms, i.e., the usual $SM$ processes such as
gluon-gluon fusion, vector-boson fusion, $t\overline{t}$ associated
production and Higgs-strahlung. For reference, we show in dashed lines
the $SM$ case (only for $h_1$), that corresponds to $\alpha =0$.

Comparing figure~\ref{xs_h1_14} to figure~\ref{xs_h1_7}, there is a
factor two enhancement passing from $\sqrt{s}=7$ TeV to $\sqrt{s}=14$
TeV centre-of-mass energy at the LHC. 

The cross-sections are a smooth function of the mixing angle $\alpha$,
so as expected every sub-channel has a cross-section that scales with
$\cos{\alpha}$ ($\sin{\alpha}$), respectively for $h_1$ ($h_2$). As a
general rule, the cross-section for $h_1$ at an angle $\alpha$ is
equal to that one of $h_2$ for $\pi /2 -\alpha$. In particular, the
maximum cross-section for $h_2$ (i.e., when $\alpha =\pi/2$) 
coincides with cross-section of $h_1$ for $\alpha =0$.

We notice that these results are in agreement with the ones that have
been discussed in
\cite{BahatTreidel:2006kx,Barger:2007im,Bhattacharyya:2007pb} in the
context of a scalar singlet extension of the $SM$, having
the latter the same Higgs production phenomenology.
Moreover, as already showed in \cite{BahatTreidel:2006kx}, also in the
minimal $B-L$ context an high value of the mixing angle could lead
to important consequences for Higgs boson discovery at the LHC: a sort
of rudimental see-saw mechanism could suppress $h_1$ production below
an observable rate at $\sqrt{s}=7$ TeV and favour just heavy Higgs
boson production, with peculiar final states clearly beyond the $SM$, or
even hide the production of both (if no more than $1$ fb$^{-1}$ of
data is accumulated). Instead, at $\sqrt{s}=14$ TeV we expect that at
least one Higgs boson will be observed, either the light one or the
heavy one, or indeed both, thus shedding light on the scalar sector of
the $B-L$ extension of the $SM$ discussed in this work. The region of
the parameter space that would allow the scalar sector to be
completely hidden, for example for $\alpha \simeq \pi/2$ and $m_{h_2}$
heavy enough to not be produced, whatever the value of $m_{h_1}$, is
experimentally excluded by precision analyses at LEP
\cite{Dawson:2009yx}.

\subsection{Non-standard production mechanisms}
\begin{figure}[!h]
  \subfloat[]{ 
  \label{xs_Zp-h1}
  \includegraphics[angle=0,width=0.48\textwidth ]{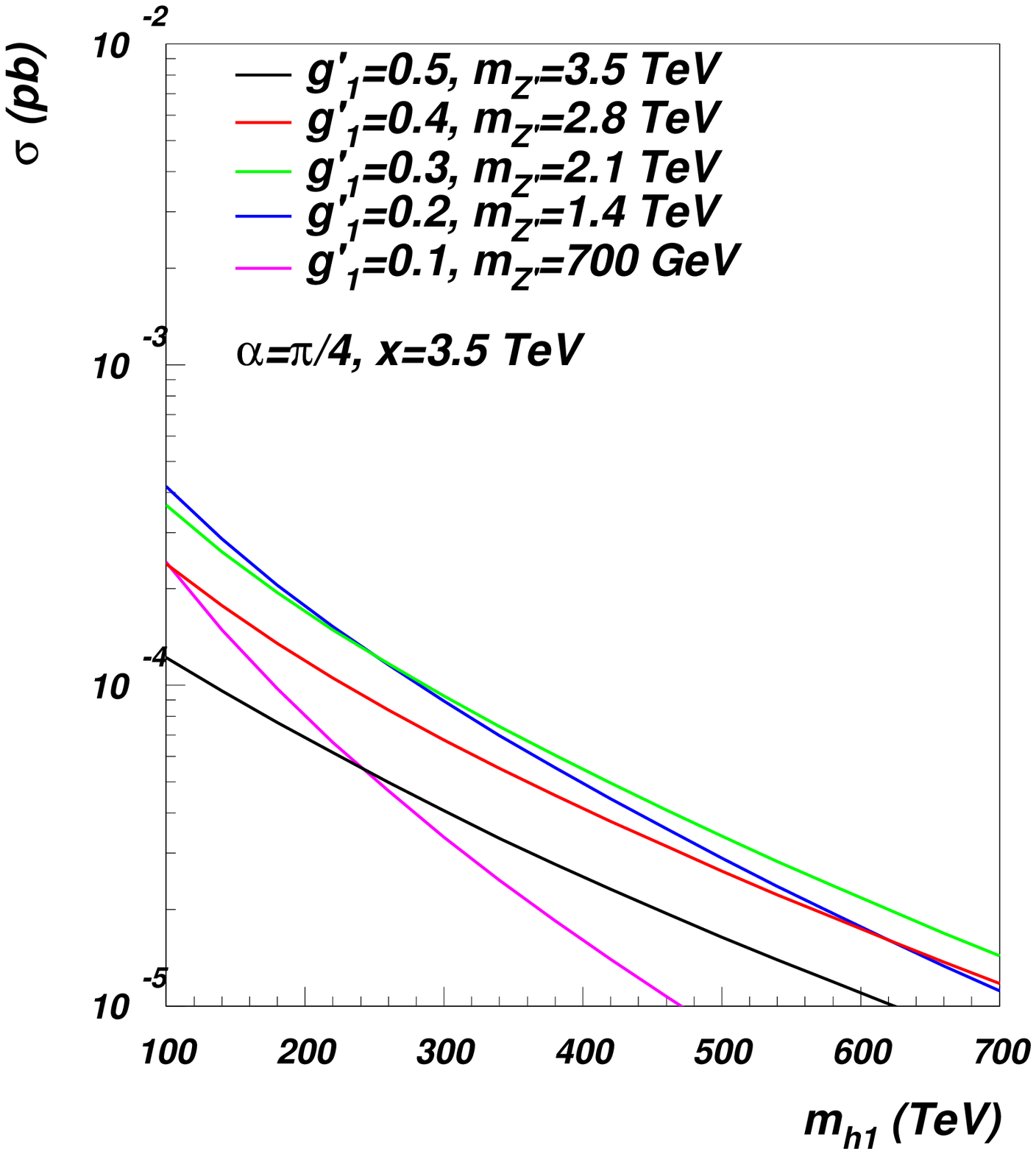}
}
  \subfloat[]{
  \label{xs_Zp-h2}
  \includegraphics[angle=0,width=0.48\textwidth ]{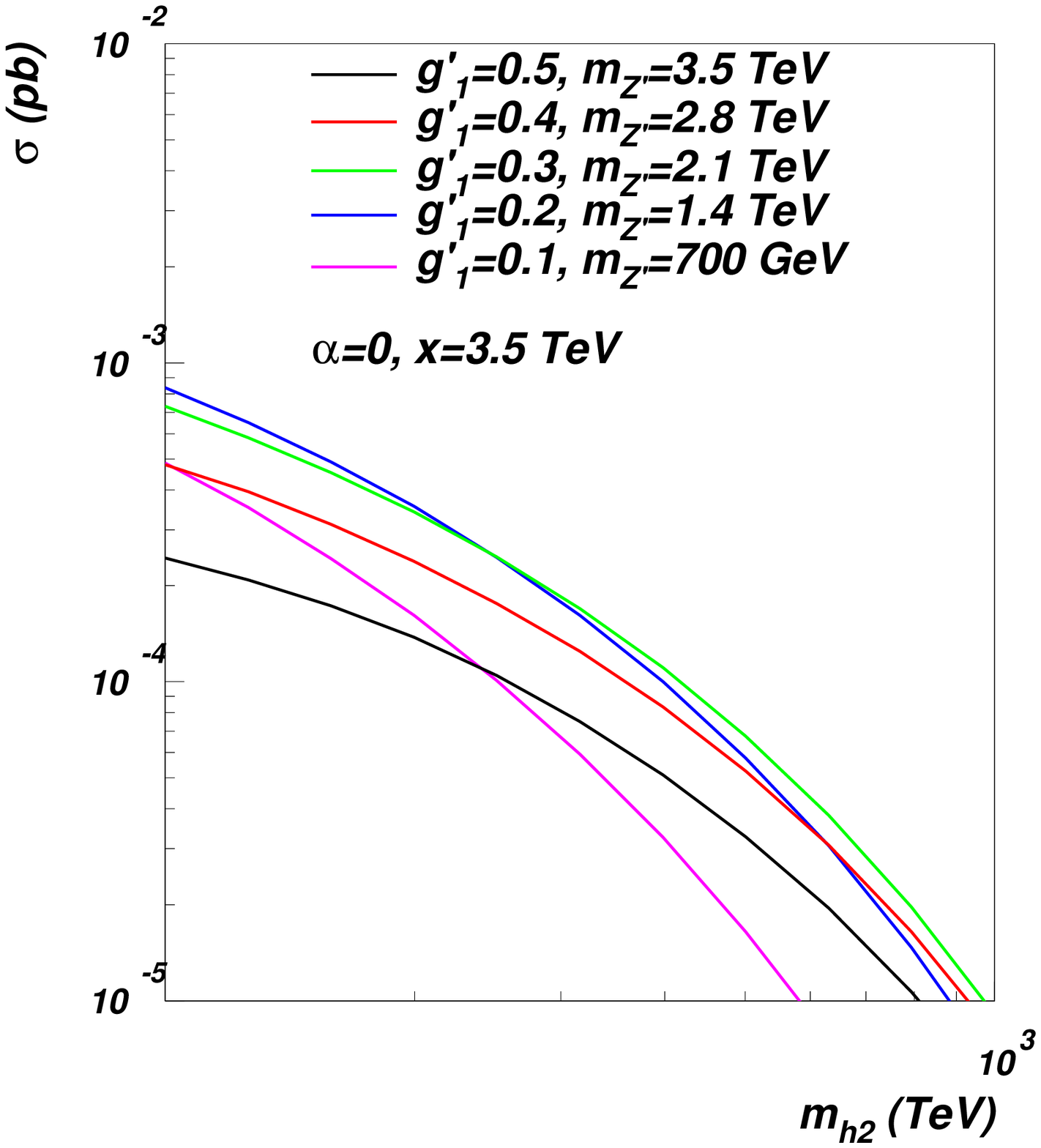}
}
  \vspace*{-0.5cm}
  \caption{\it Cross-sections in the $B-L$ model for the associated
    production with the $Z'_{B-L}$ boson (\ref{xs_Zp-h1}) of $h_1$ at
    $\alpha = \pi /4$ and (\ref{xs_Zp-h2}) of $h_2$ at $\alpha = 0$.}
  \label{strah-Xs}
\end{figure}

\begin{figure}[!h]
  \subfloat[]{
  \label{xs_h1nn}
  \includegraphics[angle=0,width=0.48\textwidth ]{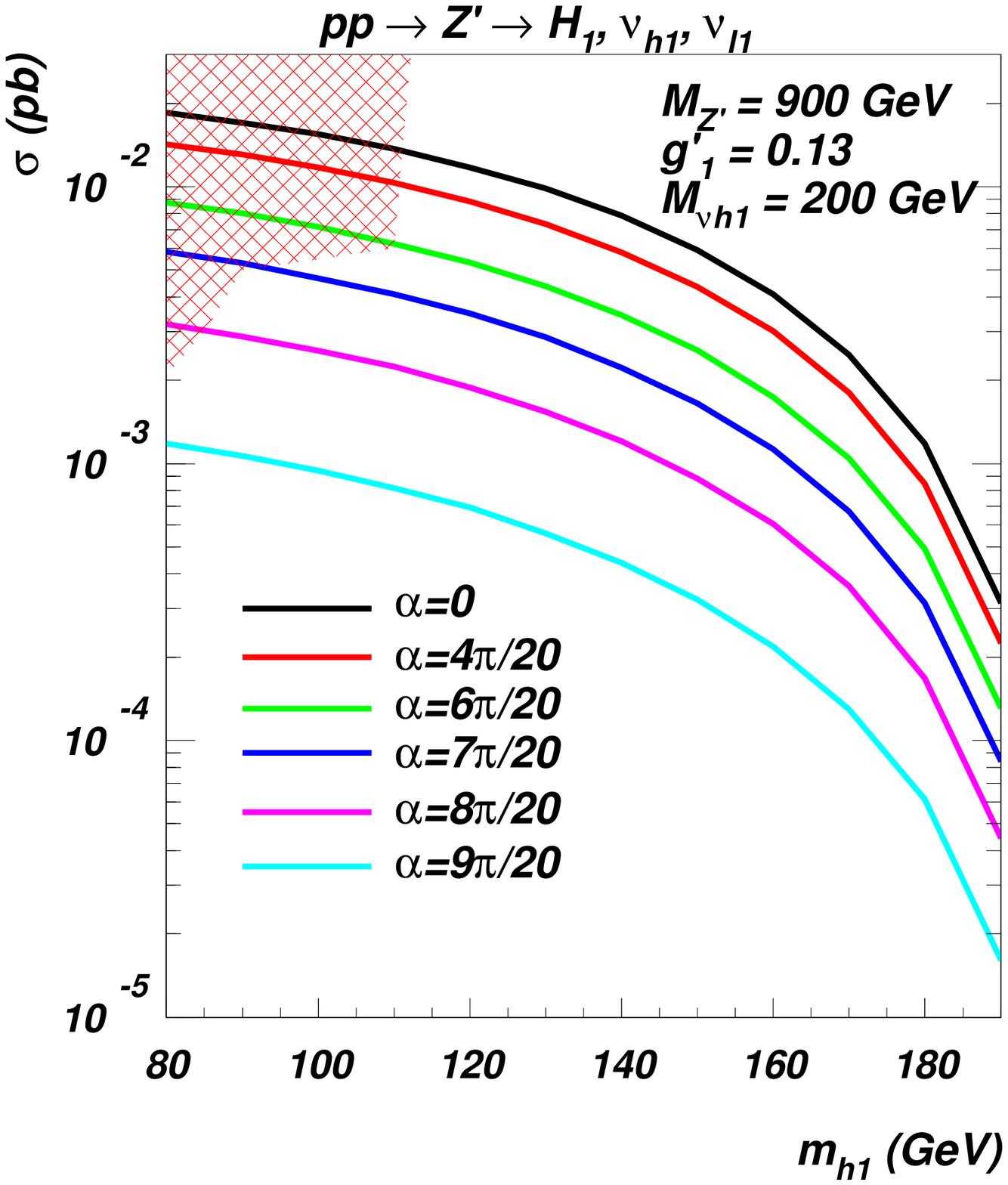}}  \\
  \subfloat[]{
  \label{xs_h1A}
  \includegraphics[angle=0,width=0.48\textwidth
  ]{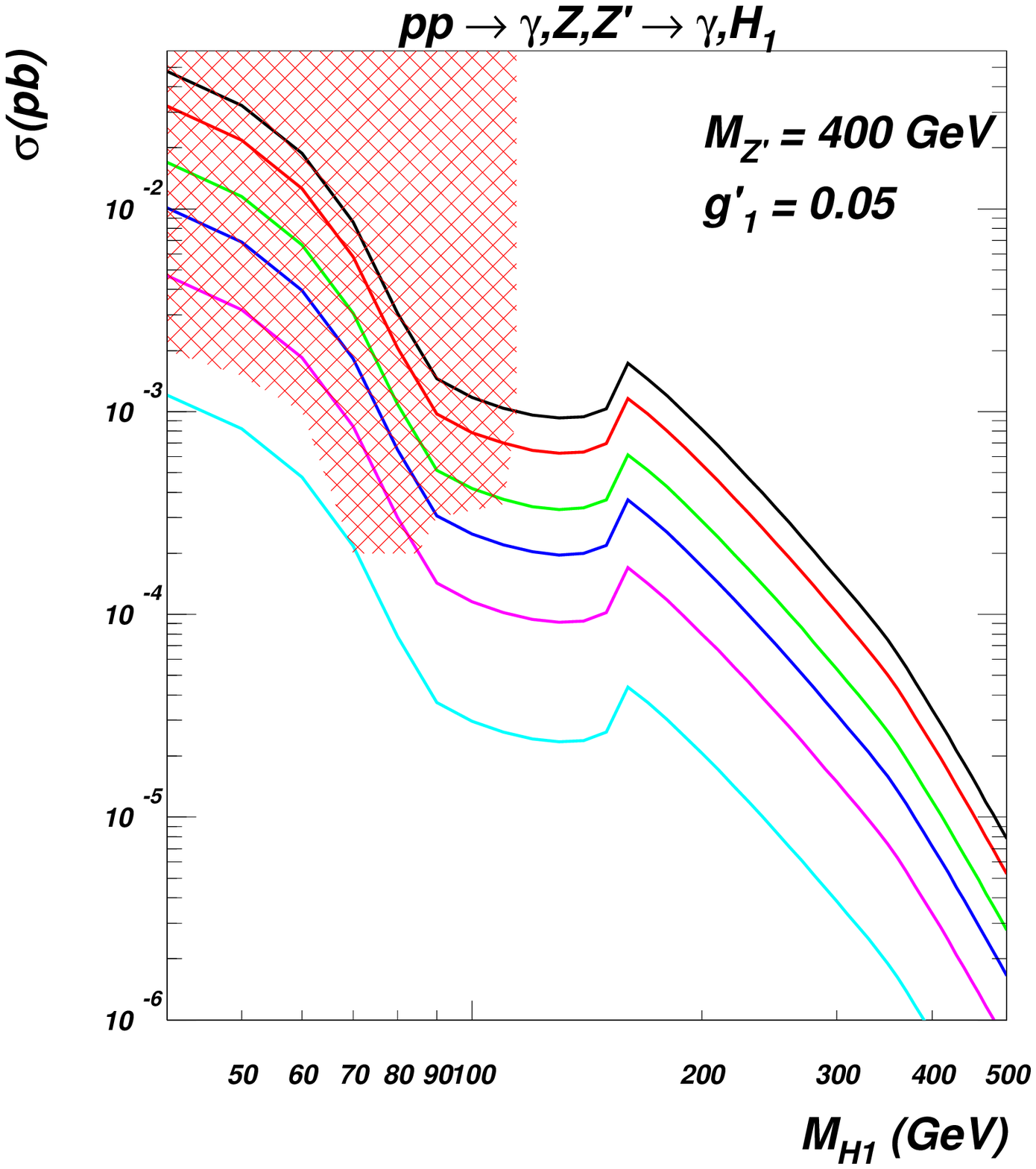}}
  \subfloat[]{
  \label{xs_h1A_VBF}
  \includegraphics[angle=0,width=0.48\textwidth ]{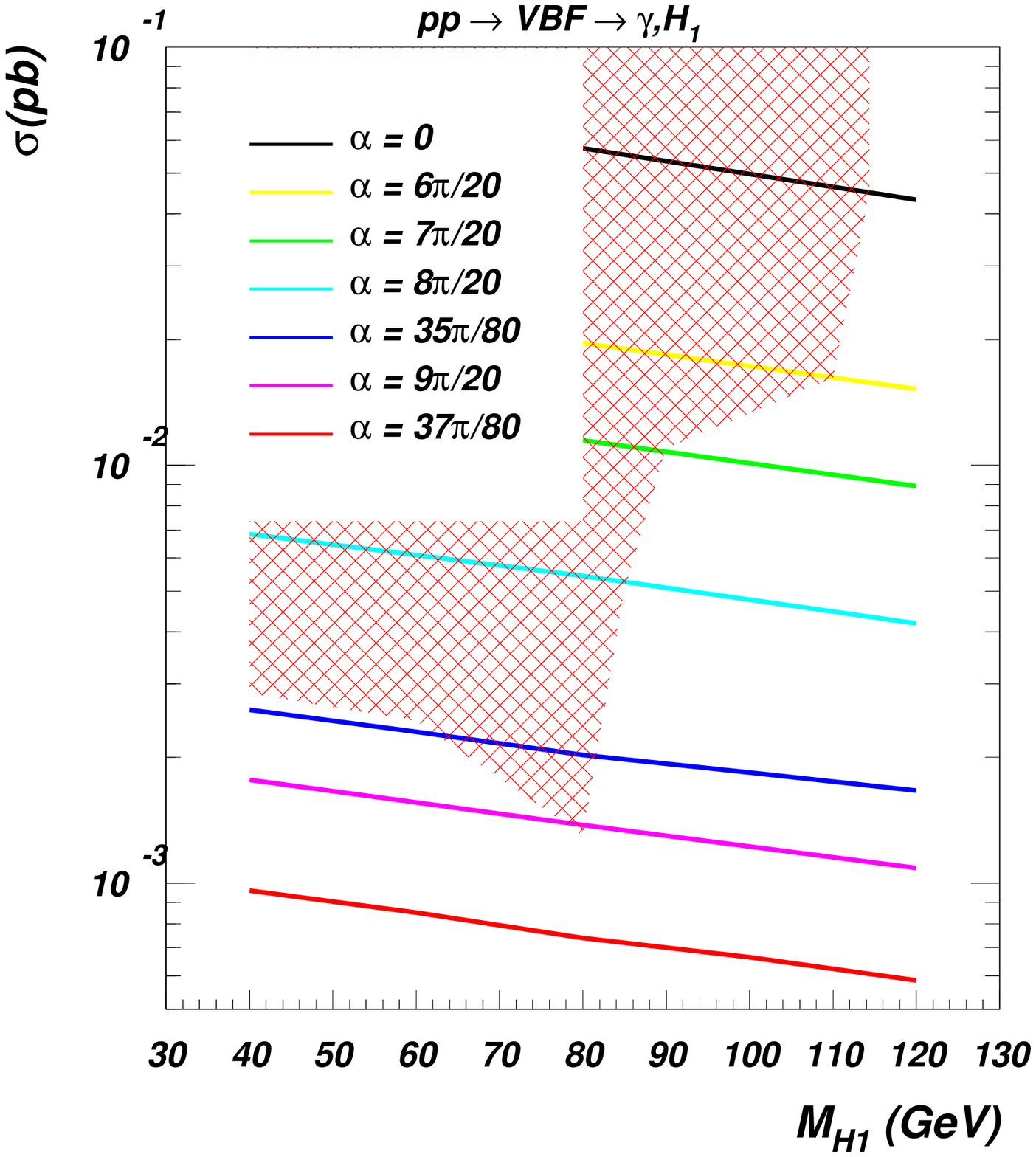}} 
  \vspace*{-0.5cm}
  \caption{\it Cross-sections in the $B-L$ model for the associated
    production of $h_1$ (\ref{xs_h1nn}) with one heavy and one light
    neutrinos, (\ref{xs_h1A}) with a photon via $\gamma$, $Z$ and $Z'$
    bosons exchange (same legend as in fig.~(\ref{xs_h1nn}) applies
    here) and (\ref{xs_h1A_VBF}) in the vector-boson fusion, all at
    $\sqrt{s}=14$ TeV. The red shading is the region excluded by LEP
    constraints \cite{Barate:2003sz}.}
  \label{non-std-Xs}
\end{figure}

All the new particles in the $B-L$ model interact with the scalar
sector, so novel production mechanisms can arise considering the
exchange of new intermediate particles. Among the new production
mechanisms, the associated production of the scalar boson with the
$Z'$ boson and the decay of a heavy neutrino into a Higgs boson are
certainly the most promising, depending on the specific masses. Notice
also that the viable parameter space, that allows a Higgs mass lighter
than the $SM$ limit of $114.4$ GeV for certain $\alpha - m_{h_2}$
configurations, enables us to investigate also production mechanisms
that in the $SM$ are subleading, as the associated production of a Higgs
boson with a photon. Figures~\ref{strah-Xs} and \ref{non-std-Xs} show
the cross-sections for the non-standard production mechanisms, for
$\sqrt{s}=14$ TeV and several values of $\alpha$.

Figures~\ref{xs_Zp-h1} and \ref{xs_Zp-h2} show the cross-sections for
associated production with the $Z'$ boson of $h_1$ and of $h_2$,
respectively, for several combinations of $Z'$ boson masses and $g'_1$
couplings. The process is
\begin{equation}
q\,\overline{q}\rightarrow Z'^{\ast} \rightarrow Z'\, h_{1(2)}\, ,
\end{equation}
and it is dominated by the $Z'$ boson's production cross-sections (see
\cite{Basso:2008iv,Basso:2010pe}). 
Although never dominant (always below $1$ fb), this channel is the
only viable mechanism to produce $h_2$ in the decoupling scenario,
i.e., $\alpha =0$.

In figure~\ref{non-std-Xs} we plot the cross-sections of the other
non-standard production mechanisms against the light Higgs mass, for
several choices of parameters (as explicitly indicated in the labels).
We superimposed the red-shadowed region in order to avoid any value of
the cross-section that has been already excluded by LEP constraints
(see \cite{Barate:2003sz}, where the relation between the reduced
coupling, in this model, is $\xi^2=\cos^2{\alpha}$), mapping each
value of the boundary cross-section as produced by the related maximum
value allowed for the 
light Higgs mass $m_{h_1}$ (at fixed mixing angle $\alpha$).

First of the showed plots is the decay of a heavy neutrino into a
Higgs boson. The whole process chain is
\begin{equation}
q\,\overline{q}\rightarrow Z' \rightarrow \nu _h\, \nu _h \rightarrow
\nu _h\, \nu _l\, h_{1(2)}\, ,
\end{equation}
and it requires to pair produce heavy neutrinos, again via the $Z'$
boson (see \cite{Basso:2008iv,Perez:2009mu} for a detailed analysis of
the $pp\rightarrow Z'\rightarrow \nu_h \nu_h$ process and other
aspects of $Z'$ and heavy 
neutrinos phenomenology in the minimal $B-L$ model). Although rather
involved, this mechanism has the advantage that
the whole decay chain can be of on-shell particles, besides the
peculiar final state of a Higgs boson and a heavy neutrino. For a
choice of the parameters that roughly maximises this mechanism
($M_{Z'}=900$ GeV, $g'_1=0.13$ and $m_{\nu _h}=200$ GeV),
figure~\ref{xs_h1nn} shows that the cross-sections for the production
of the light Higgs boson (when only one generation of heavy neutrinos
is considered) are above $10$ fb for $m_{h_1} < 130$ GeV (and small
values of $\alpha$), dropping steeply when the light Higgs boson mass
approaches the kinematical limit for the heavy neutrino to decay into
it. Assuming the transformation $\alpha \rightarrow \pi/2 -
\alpha$, the production of the heavy Higgs boson via this mechanism
shows analogous features.

Next, figures~\ref{xs_h1A} and \ref{xs_h1A_VBF} shows the associated
production of the light Higgs boson with a photon. The processes are,
respectively,
\begin{equation}\label{H1_A_GGfus}
q\,\overline{q}\rightarrow \gamma / Z / Z' \rightarrow \gamma \, h_{1}
\end{equation}
via the $SM$ neutral gauge bosons ($\gamma$ and $Z$) and the new $Z'$ boson, and
\begin{equation}\label{H1_A_VBF}
q\,q'\rightarrow \gamma \, h_{1}\, q'' \, q''' \, ,
\end{equation}
through vector-boson fusion (only $W$ and $Z$ bosons).

In the first instance, we notice that the $Z'$ sub-channel in
eq.~(\ref{H1_A_GGfus}) is always negligible,
as there is no $Z'-W-W$ interaction and the $V-h-\gamma$ effective
vertex is only via a top quark loop (an order of magnitude lower than
the $V-h-\gamma$ effective vertex via a $W$ boson loop)
\cite{Gunion:1989we}. What is relevant in these two channels is that
the light Higgs
boson mass can be considerably smaller than the LEP limit
(they are valid for the $SM$, or equivalently when $\alpha=0$ in the
$B-L$ model). Hence, the phase space factor can enhance the mechanism
of eq.~(\ref{H1_A_GGfus}) for small masses, up to the level of $1$ fb
for $m_{h_1} < 60$ GeV (and suitable values of the mixing angle
$\alpha$, depending on the experimental and theoretical limits, see
Refs.~\cite{Basso:2010jt,Basso:2010jm} for a complete tratement of the
allowed parameter space of the Higgs sector of the minimal $B-L$
problem). Moreover, it has recently been
observed that the associated production with a photon in the 
vector-boson fusion channel could be useful for low Higgs boson masses
to trigger
events in which the Higgs boson decays into $b$-quark pairs
\cite{Asner:2010ve}. Complementary to that, the process in
eq.~(\ref{H1_A_GGfus}) can also be of similar interest, with the
advantage that the photon will always be back-to-back relative to the
$b$-quark pair. For comparison, figures~\ref{xs_h1A} and
\ref{xs_h1A_VBF} show the cross-section for these
processes\footnote{In order to produce figure~\ref{xs_h1A_VBF}, we
  included the following cuts: $P_t ^{\gamma,\mbox{jet}} > 15$ GeV,
  $|\eta ^{\gamma}| < 3$ and $|\eta ^{\mbox{jet}}|~<~5.5$, where
  ``jet'' refers to the actual final state, though we use partons here to emulate it 
  \cite{Asner:2010ve}.}. Certainly, for a ${h_1}$ boson heavier than
the $SM$ limit, vector-boson fusion is the dominant process for
associated production of $h_1$ with a photon, and this is also true
for $m_{h_1} > 60$~GeV. However, for light Higgs boson masses lower
than $60$~GeV, the two mechanisms of eqs.~(\ref{H1_A_GGfus}) and
(\ref{H1_A_VBF}) become equally competitive, up to the level of
$\mathcal{O}(1)$~fb each, for suitable values of the mixing angle
$\alpha$.

\subsection{Branching ratios and total widths}
\begin{figure}[!h]
  \subfloat[]{ 
  \label{BR_h1}
  \includegraphics[angle=0,width=0.48\textwidth ]{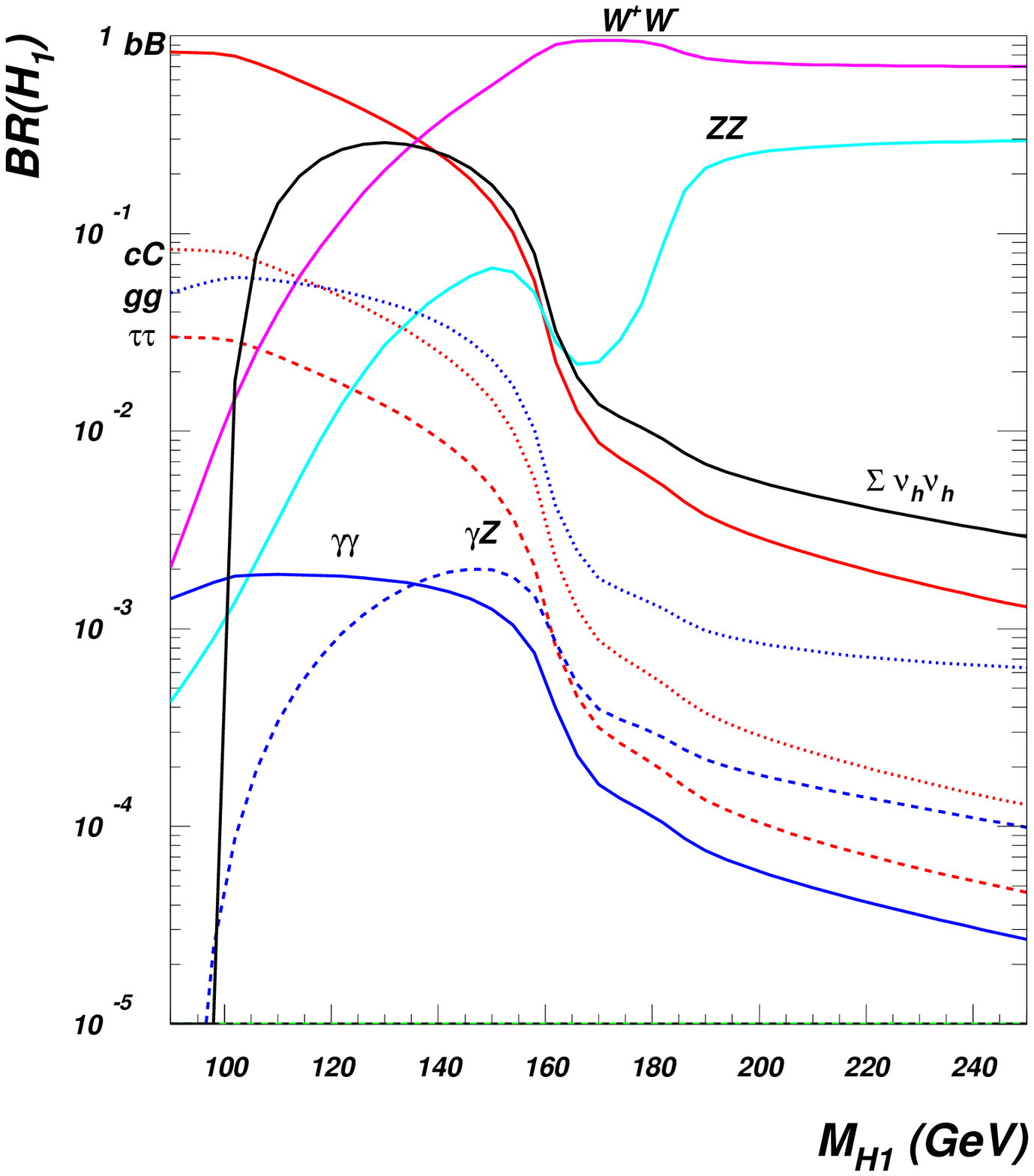}}
  \subfloat[]{
  \label{BR_h2}
  \includegraphics[angle=0,width=0.48\textwidth ]{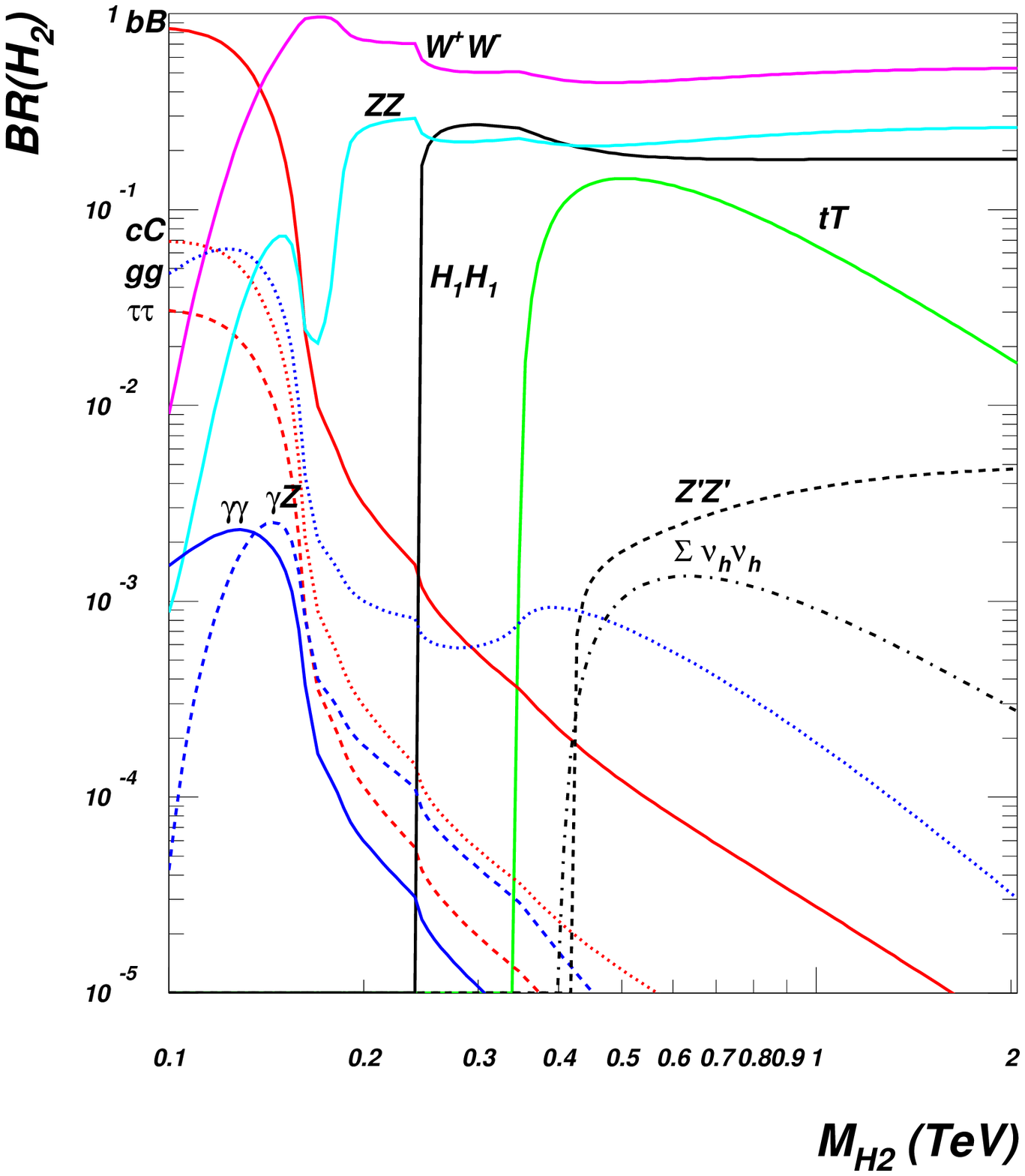}} \\
  \subfloat[]{ 
  \label{H1_TW}
  \includegraphics[angle=0,width=0.48\textwidth ]{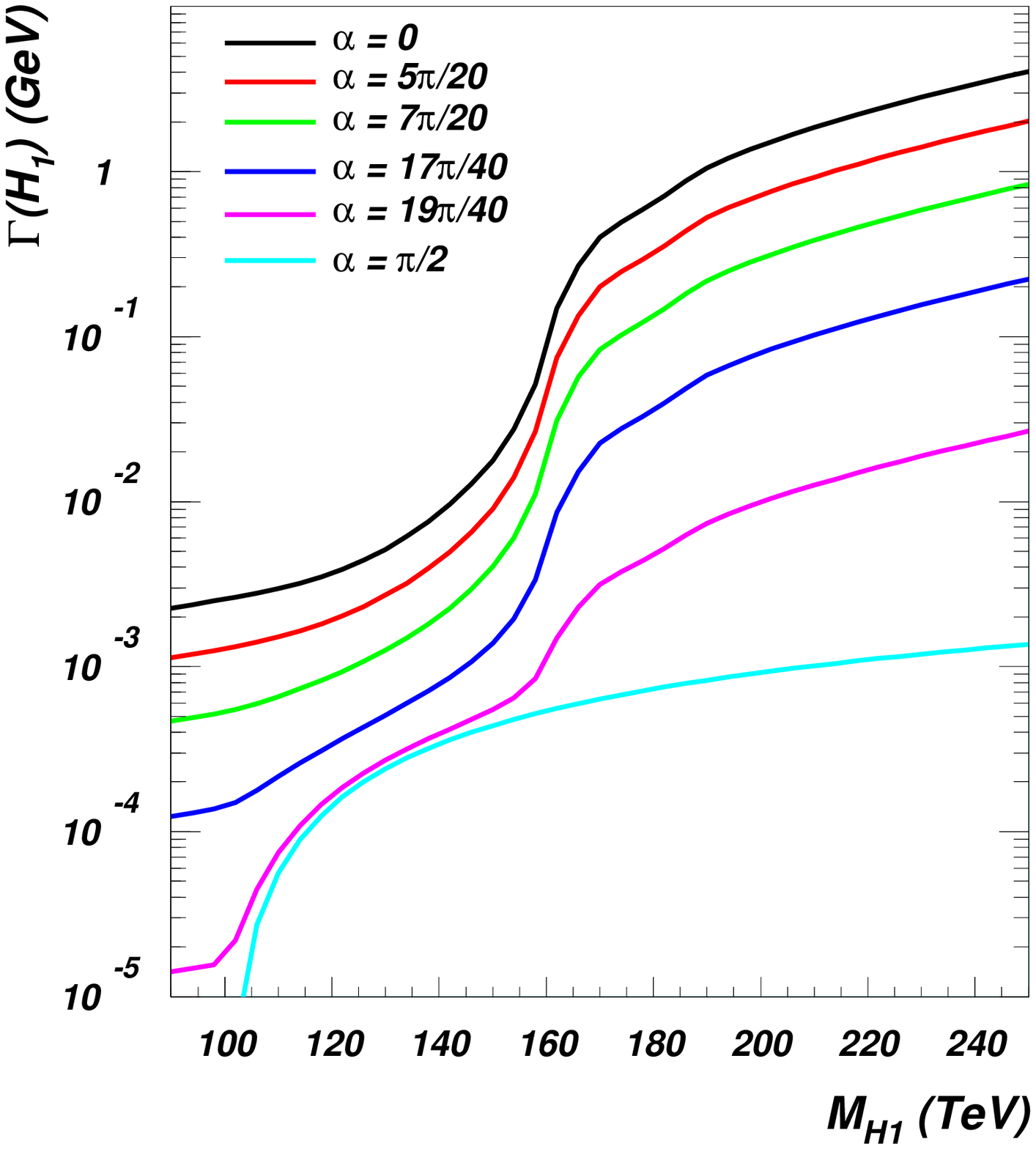}}
  \subfloat[]{
  \label{H2_TW}
  \includegraphics[angle=0,width=0.48\textwidth ]{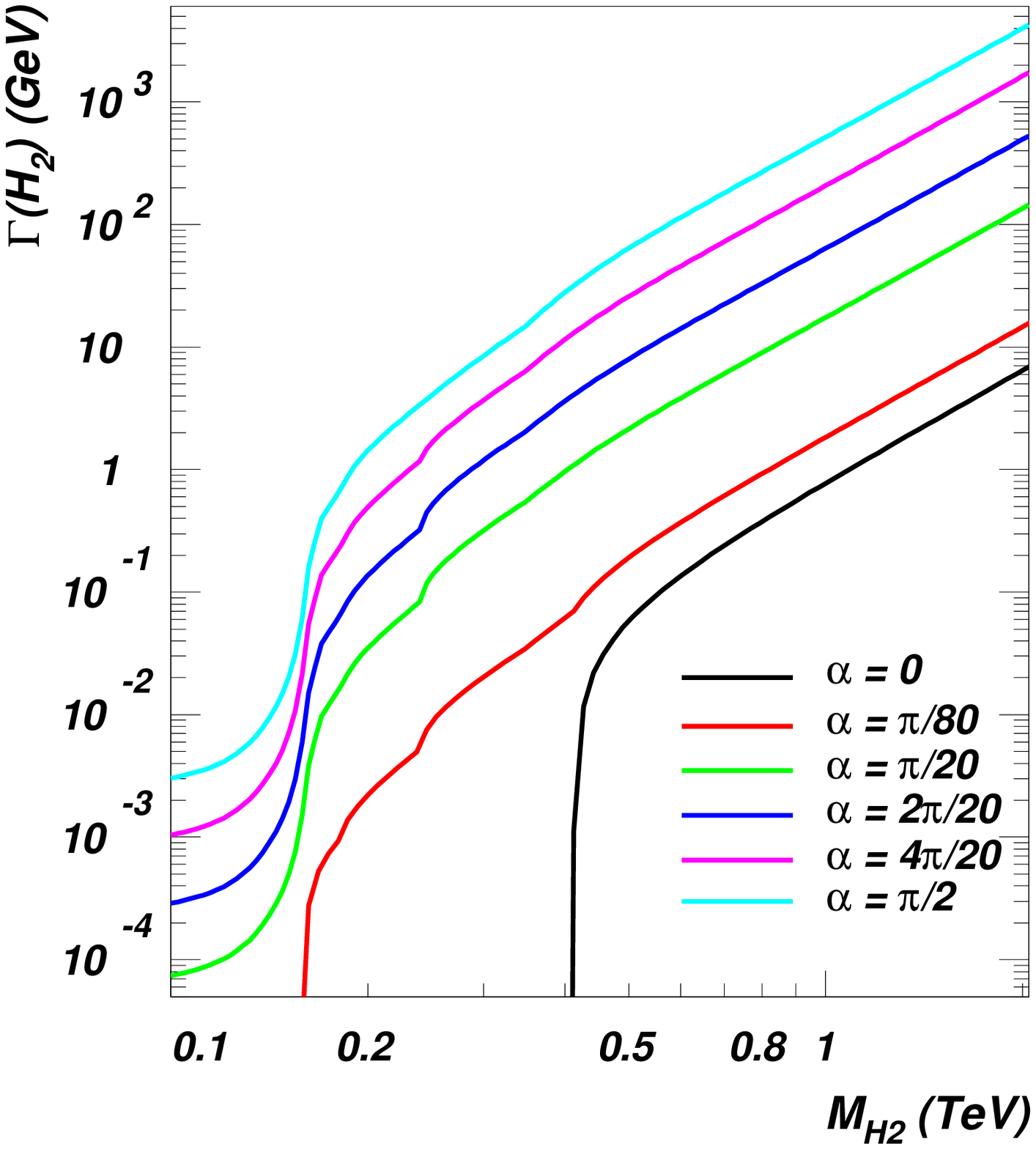}} 
  \vspace*{-0.5cm}
  \caption{\it (\ref{BR_h1}) Branching ratios for $h_1$ for $\alpha =2\pi /5$ and $m_{\nu_h}=50$ GeV and (\ref{H1_TW}) $h_1$ total width for a choice of mixing angles and (\ref{BR_h2}) BRs for $h_2$ for $\alpha =3\pi /20$ and $m_{h_1}=120$ GeV, $M_{Z'}=210$ GeV and $m_{\nu_h}=200$ GeV and (\ref{H2_TW}) $h_2$ total width for a choice of mixing angles.}
  \label{Brs}
\end{figure}

Moving to the Higgs boson decays, figure~\ref{Brs} shows the BRs for both the Higgs bosons, $h_1$ and $h_2$, respectively. Only the two-body decay channels are shown here.

Regarding the light Higgs boson, the only new particle it can decay into is the heavy neutrino (we consider a very light $Z'$ boson unlikely and unnatural), if the channel is kinematically open. In figure~\ref{BR_h1} we show this case, for a small heavy neutrino mass, i.e., $m_{\nu_h}=50$ GeV, and we see that the relative BR of this channel can be rather important, as the decay into 
$b$-quark pairs or into $W$ boson pairs, in the range of masses $110$ GeV $\leq m_{h_1} \leq 150$ GeV. Such range happens to be critical in the $SM$ since here the $SM$ Higgs boson passes from decaying dominantly into 
$b$-quark pairs to a region in masses in which the decay into $W$ boson pairs is the prevailing one. These two decay channels have completely different signatures and discovery methods/powers. The fact that the signal of the Higgs boson decaying into 
$b$-quark pairs is many orders of magnitude below the natural QCD background, spoils its sensitivity. In the case of the $B-L$ model, the decay into heavy neutrino pairs is therefore phenomenologically very important, besides being an interesting feature of the $B-L$ model if $m_{\nu_h} < M_W$, as it allows multileptons signatures of the light Higgs boson. Among them, there is the decay of the Higgs boson into $3\ell$, $2j$ and $\met$ (that we have already studied for the $Z'$ case in Ref.\cite{Basso:2008iv} and that will be reported 
upon separately for the Higgs boson case \cite{bbmp}), into $4\ell$ and $\met$ (as, again, already studied for 
the $Z'$ case in Ref.~\cite{Huitu:2008gf}) or into $4\ell$ and $2j$ (as already studied, when $\ell = \mu$, in the $4^{th}$ family extension of the $SM$ \cite{CuhadarDonszelmann:2008jp}). All these peculiar signatures allow the Higgs boson signal to be studied in channels much cleaner than the decay 
into $b$-quark pairs.

\begin{figure}[!h]
  \subfloat[]{ 
  \label{H2_BR-a_H1}
  \includegraphics[angle=0,width=0.48\textwidth ]{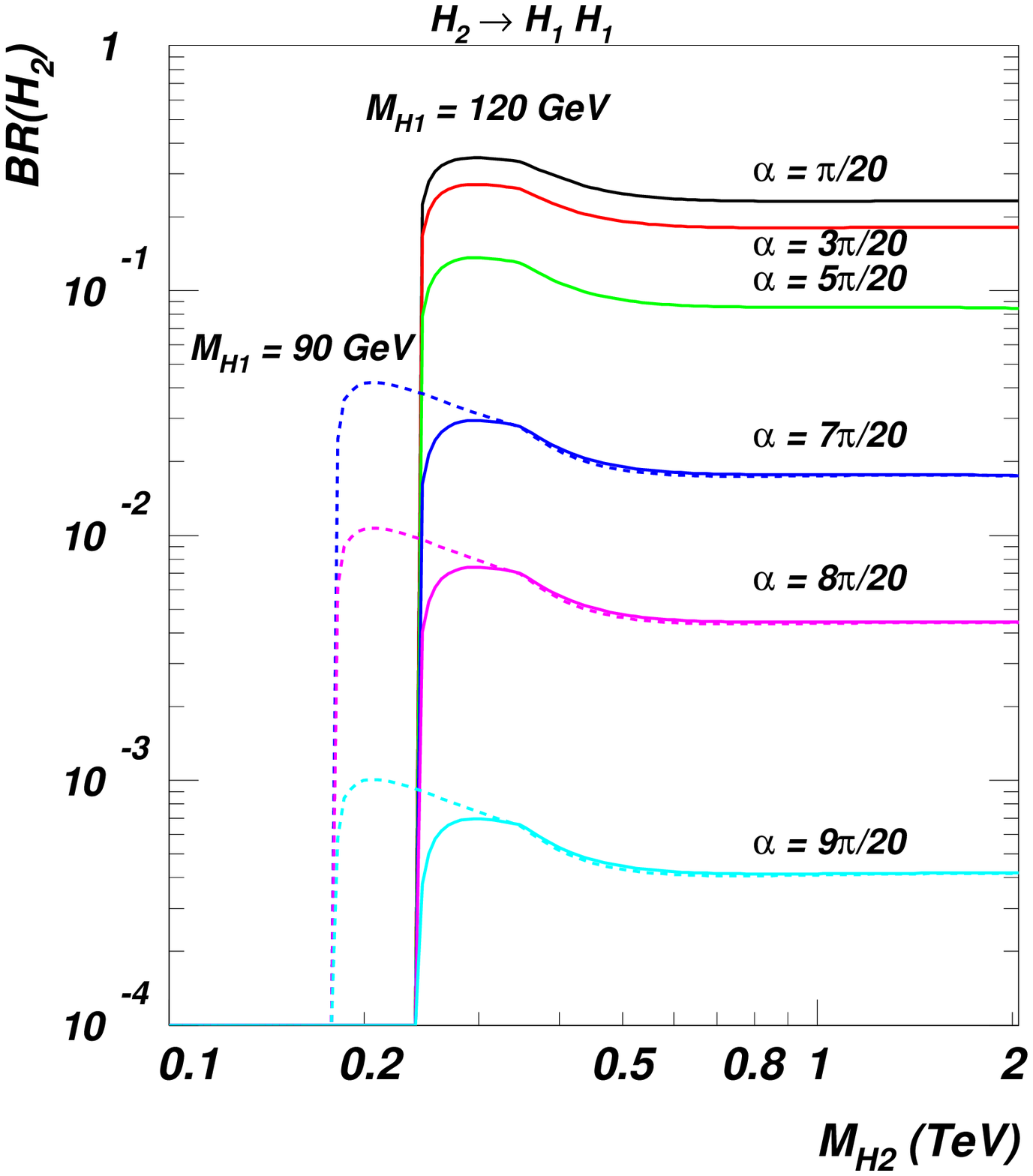}
}\\
  \subfloat[]{
  \label{H2_BR-a_Hnu}
  \includegraphics[angle=0,width=0.48\textwidth ]{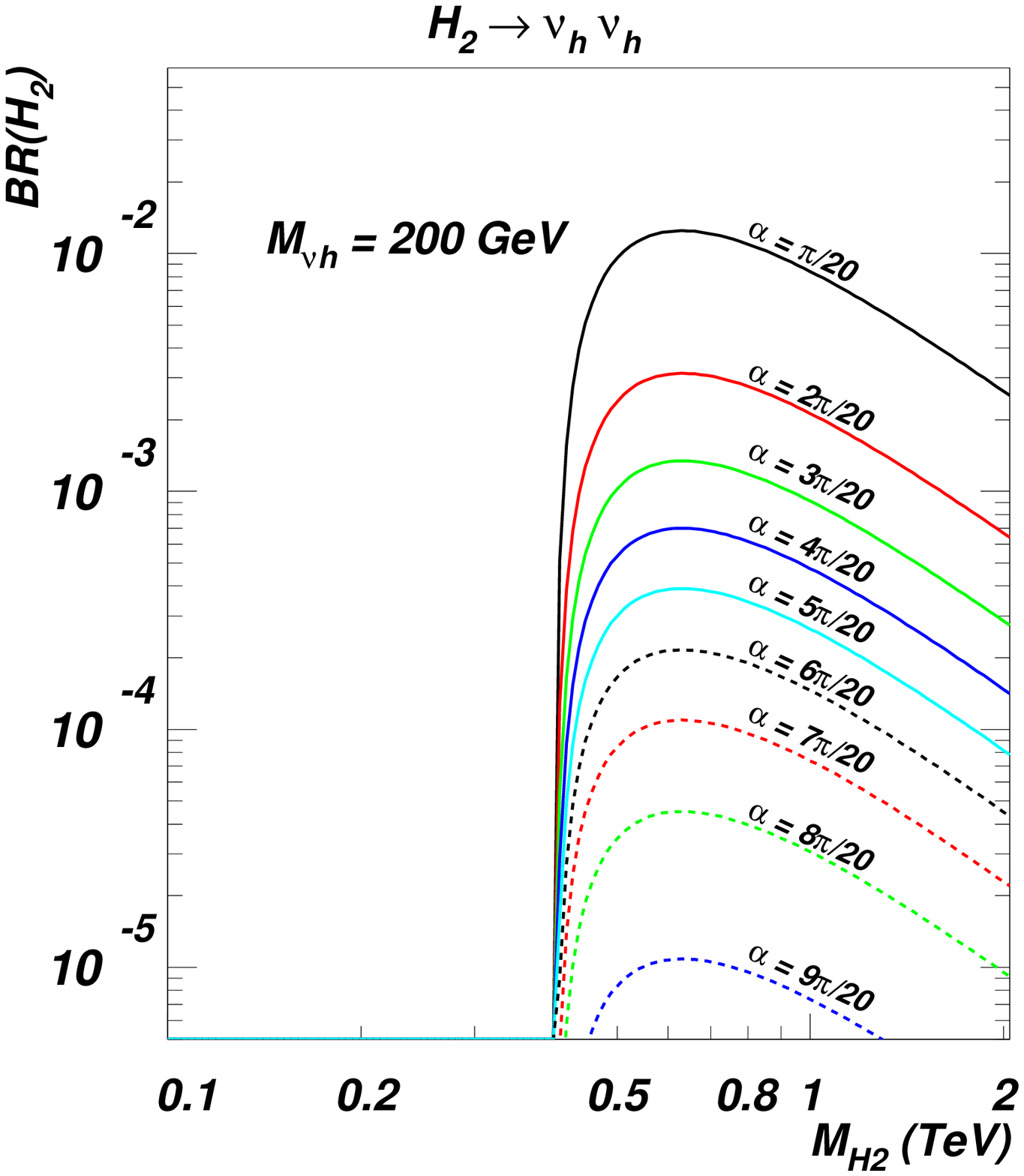}}  
  \subfloat[]{
  \label{H2_BR-a_Zp}
  \includegraphics[angle=0,width=0.48\textwidth ]{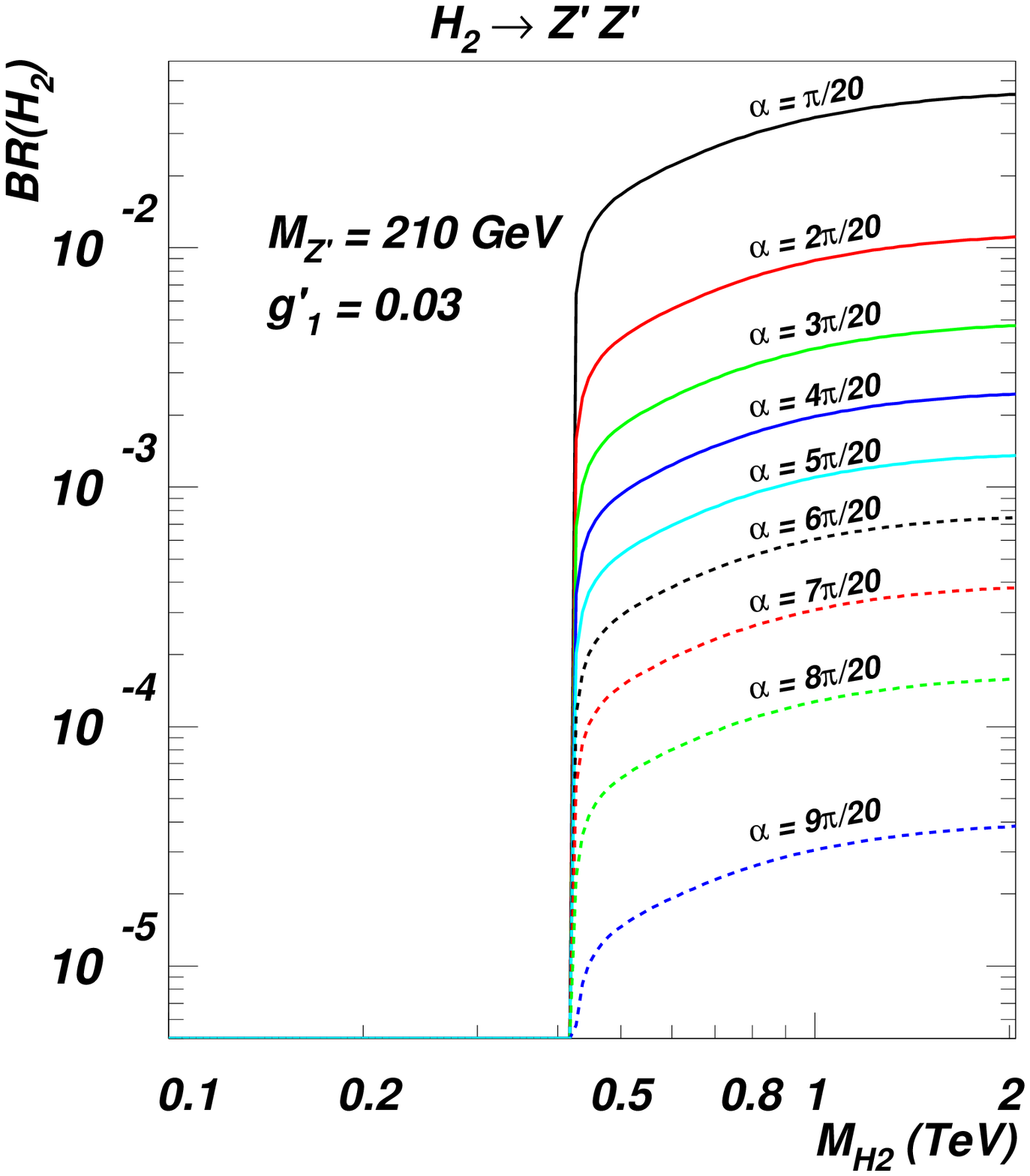}}  
  \vspace*{-0.5cm}
  \caption{\it Dependence on the mixing angle $\alpha$ of (\ref{H2_BR-a_H1}) $BR(h_2\rightarrow h_1\, h_1)$,  of (\ref{H2_BR-a_Hnu}) $BR(h_2\rightarrow \nu _h\, \nu _h)$ and of (\ref{H2_BR-a_Zp}) $BR(h_2\rightarrow Z'\, Z')$.}
  \label{Br-alpha}
\end{figure}

In the case of the heavy Higgs boson, further decay channels are possible in the $B-L$ model, if kinematically open. The heavy Higgs boson can decay in pairs of the light Higgs boson ($h_2 \rightarrow h_1\, h_1$) or even in triplets ($h_2 \rightarrow h_1\, h_1\, h_1$), in pairs of heavy neutrinos and $Z'$ bosons. Even for a small value of the angle, figure~\ref{BR_h2} shows that the decay of a heavy Higgs boson into pairs of the light one can be quite sizeable, at the level of the decay into $SM$ $Z$ bosons for $m_{h_1} = 120$ GeV . It is important to note that this channel does not have a simple dependence on the mixing angle $\alpha$, as we can see in figure~\ref{Br-alpha}.

The BRs of the heavy Higgs boson decaying into $Z'$ boson pairs and heavy neutrino pairs decrease as the mixing angle 
increases, getting to their maxima (comparable to the $W$ and $Z$ ones) for a vanishing $\alpha$, for which the production 
cross-section is however negligible. As usual, and also clear from figure~\ref{BR_h2}, the decay of the heavy Higgs boson into gauge bosons (the $Z'$ boson) is always bigger than the decay into pairs of fermions (the heavy neutrinos, even when summed over the generations as plotted), when they have comparable masses (here, $M_{Z'}=210$ GeV and $m_{\nu _h} = 200$ GeV).

The other standard decays of both the light and the heavy Higgs bosons are not modified substantially in the $B-L$ model (i.e., the Higgs boson to $W$ boson pairs is always dominant when kinematically open, while before that
the decay into $b$-quarks is the prevailing one; further, radiative decays, such as Higgs boson decays into 
pairs of photons, peak at around $120$ GeV, etc.). Only when other new channels open, the standard decay channels 
alter accordingly. This rather common picture could be altered when the mixing angle $\alpha$ 
approaches $\pi/2$, but such situation is phenomenologically not viable \cite{Dawson:2009yx}.

Figures~\ref{H1_TW} and \ref{H2_TW} show the total widths for $h_1$ and $h_2$, respectively. In the first case, few thresholds are clearly recognisable, as the heavy neutrino one at $100$ GeV (for angles very close to $\pi /2$ only), the $W$ and the $Z$ ones. Over the mass range considered ($90$ GeV $< m_{h_1} < 250$ GeV, the particle's width )is very small until the $W$ threshold, less than $1-10$ MeV, rising steeply to few GeV for higher $h_1$ masses and small angles (i.e., for a $SM$-like light Higgs boson). As we increase the mixing angle, the couplings of the light Higgs boson to $SM$ particles is reduced, as so its total width.

On the contrary, as we increase $\alpha$, the $h_2$ total width increases, as clear from figure~\ref{H2_TW}. Also in this case, few thresholds are recognisable, as the usual $W$ and $Z$ gauge boson ones, the light Higgs boson one (at $240$ GeV) and the 
$t$-quark one (only for big angles, i.e., when $h_2$ is the $SM$-like Higgs boson). When the mixing angle is small, the $h_2$ total width stays below $1$ GeV all the way up to $m_{h_2} \sim 300\div 500$ GeV, rising as the mass increases towards values for which $\Gamma_{h_2} \sim m_{h_2}\sim 1$ TeV and $h_2$ loses the meaning of 
resonant state, only for angles very close to $\pi /2$. Instead, if the angle is small, i.e., less than $\pi /10$, the ratio of width over mass is less than $10 \%$ and the heavy Higgs boson is a well defined particle. In the decoupling regime, i.e., when $\alpha =0$, the only particles $h_2$ couples to are the $Z'$ and the heavy neutrinos. The width is therefore dominated by the decay into them and is tiny, as clear from figure~\ref{H2_TW}.

As already mentioned, figure~\ref{Br-alpha} shows the dependence on the mixing angle $\alpha$ of the 
BRs of $h_2$ into pairs of non-$SM$ particles. In particular, we consider the decays $h_2 \rightarrow h_1\, h_1$ (for two different $h_1$ masses, $m_{h_1}=90$ GeV and $m_{h_1}=120$ GeV, only for the allowed values of $\alpha$), $h_2 \rightarrow \nu _h\, \nu _h$ and $h_2 \rightarrow Z'\, Z'$ (not influenced by $m_{h_1}$). As discussed in section~\ref{Sec:Analysis_details}, the interaction of the heavy Higgs boson with $SM$ (or non-$SM$) particles has an overall $\sin{\alpha}$ (or $\cos{\alpha}$, respectively)
dependence. Nonetheless, the BRs in figure~\ref{Br-alpha} depend also on the total width, that for  $\alpha > \pi/4$ is dominated by the $h_2 \rightarrow W^+ W^-$ decay. Hence, when the angle assumes big values, the angle dependence of the $h_2$ BRs into heavy neutrino pairs and into $Z'$ boson pairs follow a simple $\cot{\alpha}$ behaviour. Regarding $h_2 \rightarrow h_1\, h_1$, its BR is complicated by the fact that the contribution of this process to the total width is not negligible when the mixing angle is small, i.e., $\alpha < \pi/4$. In general, this channel vanishes when $\alpha \rightarrow 0$, and it gets to its maximum, of around $10\% \div 30 \%$ of the total width, as $\alpha$ takes a non-trivial value, being almost constant with the angle if it is small enough.

\begin{figure}[!h]
  \subfloat[]{
  \label{H2-Br-3H1}
  \includegraphics[angle=0,width=0.48\textwidth ]{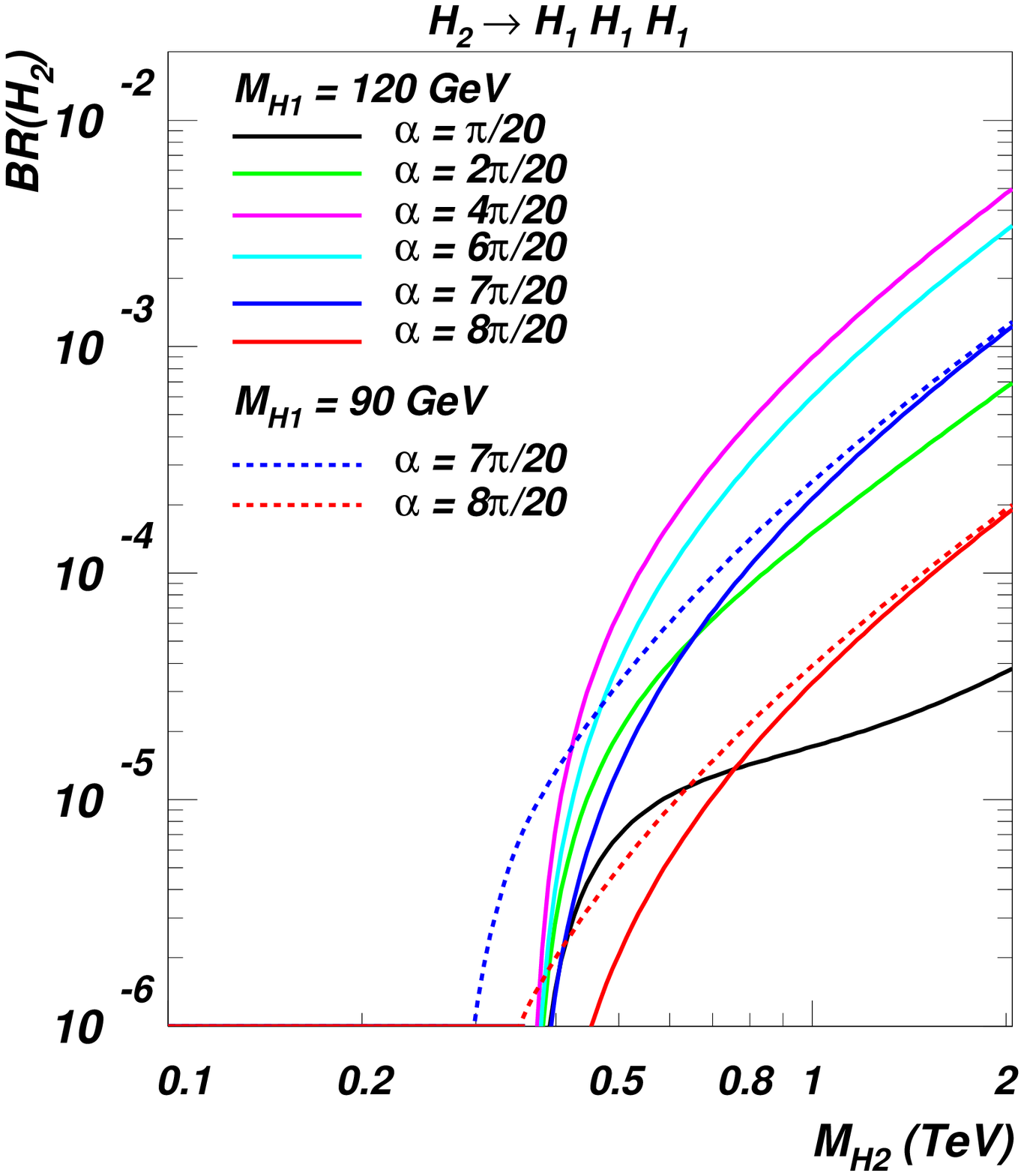}}\\
  \subfloat[]{
  \label{H2-Br-2Vh1}
  \includegraphics[angle=0,width=0.48\textwidth ]{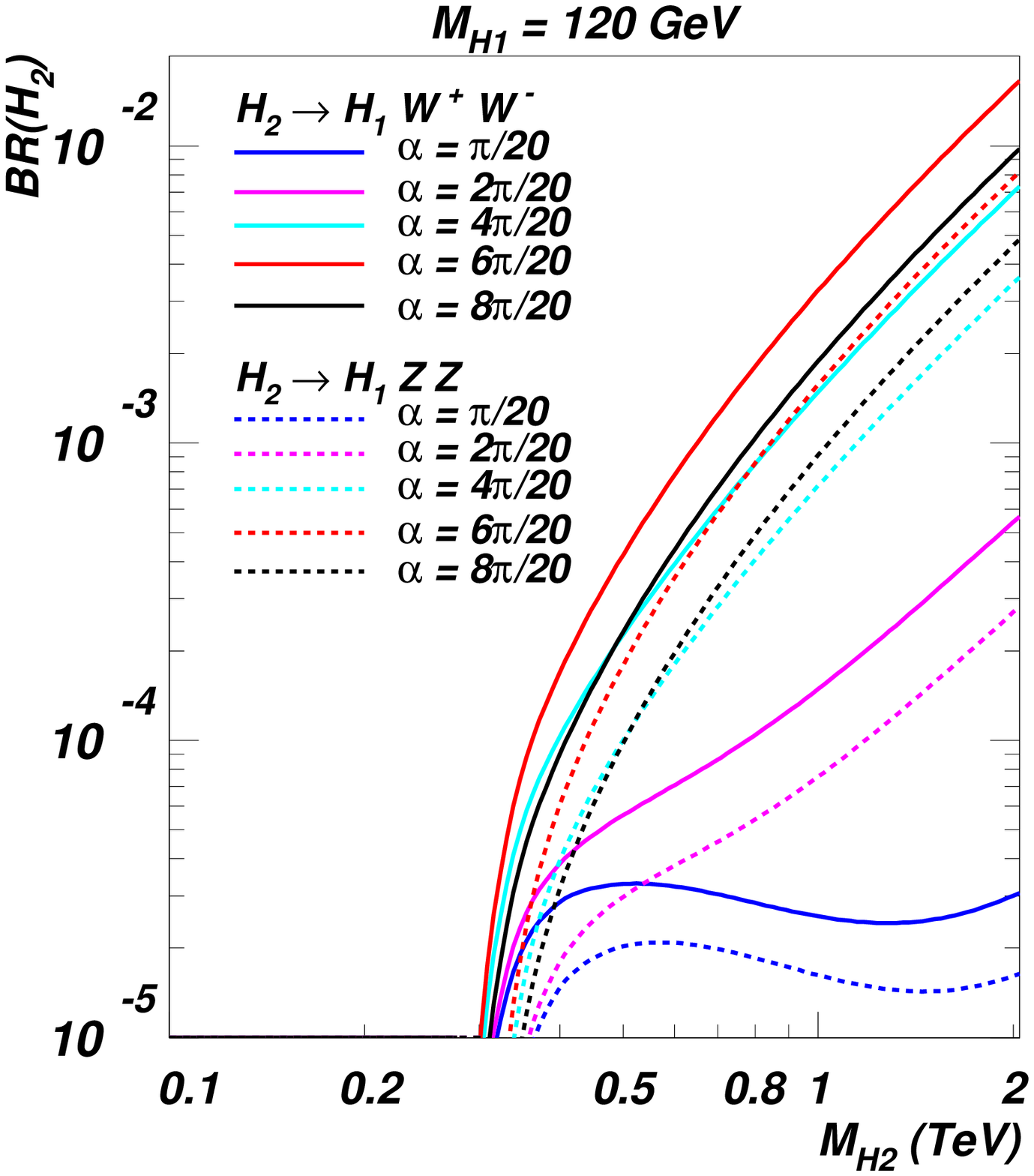}}
  \subfloat[]{
  \label{H2-Br-2Vh1-90}
  \includegraphics[angle=0,width=0.48\textwidth ]{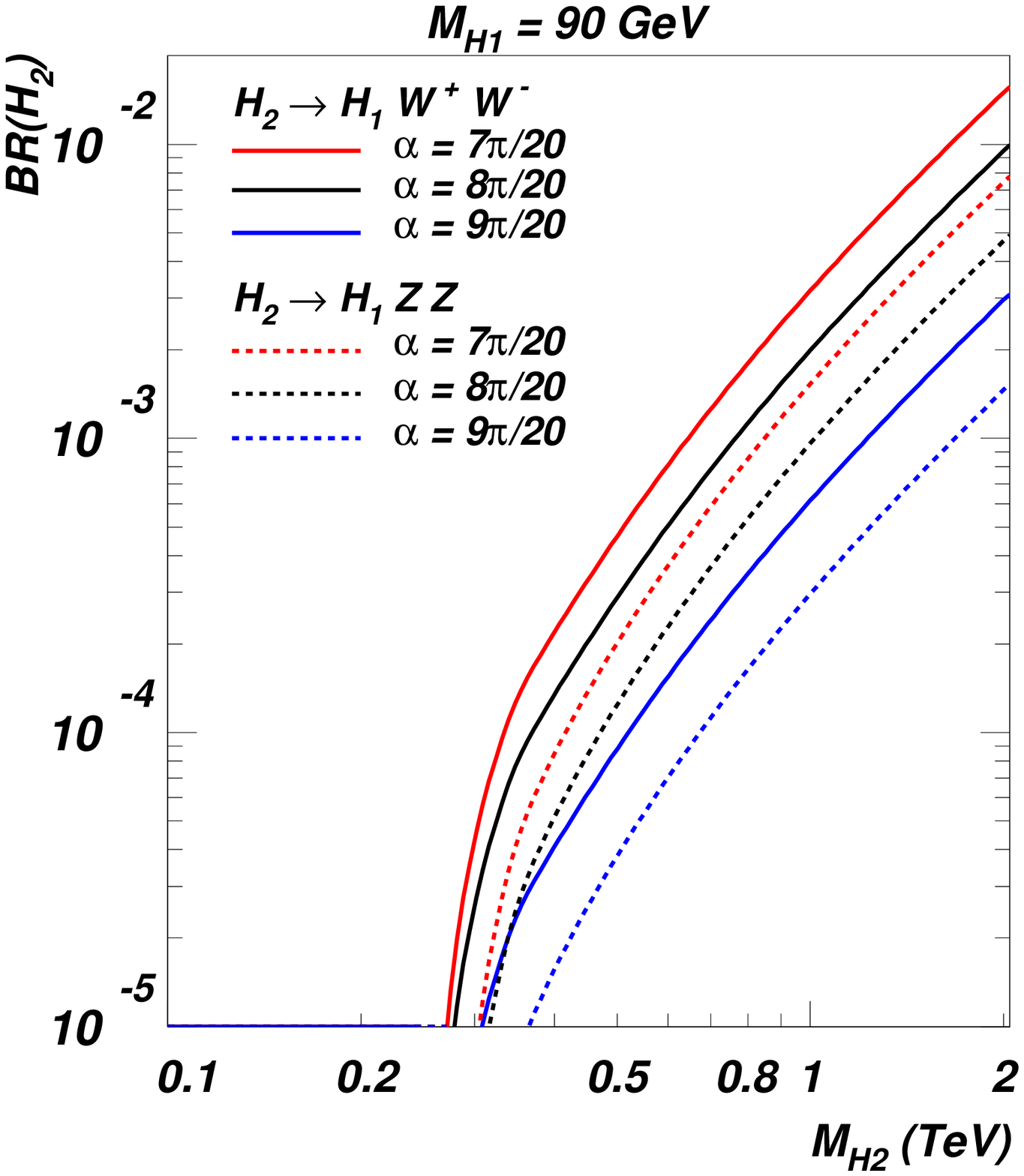}}
  \vspace*{-0.5cm}
  \caption{\it Dependence on the mixing angle $\alpha$ of the three body decays (\ref{H2-Br-3H1}) BR$(h_2\rightarrow h_1\, h_1\, h_1)$ and  (\ref{H2-Br-2Vh1}) B$R(h_2\rightarrow h_1\, V\, V)$ ($V=W^\pm,Z$) for $m_{h_1} = 120$ GeV and (\ref{H2-Br-2Vh1-90}) for $m_{h_1} = 90$ GeV, respectively.}
  \label{Br-3b}
\end{figure}

The heavy Higgs boson can be relatively massive and the tree-level three-body decays are interesting decay modes too. Besides being clear $BSM$ signatures, they are crucial to test the theory behind the observation of any scalar particle: its self-interactions and the quartic interactions with the vector bosons could be tested directly in these decay modes. In the $B-L$ model with no $Z-Z'$ mixing, the quartic interactions that can be tested as $h_2$ decay modes, if the respective channels are kinematically open, are: $h_2 \rightarrow h_1\, h_1\, h_1$, $h_2 \rightarrow h_1\, W^+\, W^-$ and $h_2 \rightarrow h_1\, Z\, Z$, as shown in figure~\ref{Br-3b}, again for $m_{h_1} = 90$~GeV and $120$~GeV. Although possible, $h_2 \rightarrow h_1\, Z'\, Z'$ is negligible always, even if the $Z'$ boson is light enough to allow the decay. For $M_{Z'}=210$ GeV, BR$(h_2 \rightarrow h_1\, Z'\, Z') \lesssim 10^{-5}$ for $m_{h_2} < 2$ TeV.

The BRs for both the $h_2 \rightarrow h_1\, h_1\, h_1$ and the $h_2 \rightarrow h_1\, V\, V$ ($V=W^\pm,\, Z$) channels are maximised roughly when the mixing between the two scalars is maximum, i.e., when $\alpha \sim \pi /4$, regardless of $m_{h_1}$. The former channel, that is interesting because would produce three light Higgs bosons simultaneously, can contribute at most at $10^{-3}$ of the total width for $h_2$, as we are neglecting values of $m_{h_2}$ and $\alpha$ for which $\Gamma_{h_2} \sim m_{h_2}$ (see figure~\ref{H2_TW}). For instance, for $m_{h_2} = 800$ GeV, $\alpha$ needs to be less than $\pi /5$ to have a reasonable small width-over-mass ratio ($\sim 10\%$), and BR$(h_2 \rightarrow h_1\, h_1\, h_1) \leq 0.6\cdot 10^{-3}$. The situation is similar for the latter channel, involving pairs of $SM$ gauge bosons. Again, for $m_{h_2} = 800$ GeV and $\alpha = \pi /5$,  BR$(h_2 \rightarrow h_1\, W^+\, W^-)$ = $2$ BR$(h_2 \rightarrow h_1\, Z\, Z) = 10^{-3}$ for $m_{h_1} = 120$ GeV. For $m_{h_1} = 90$ GeV, the mixing angle is constrained to be bigger than $7\pi /20$. For these values and the same $m_{h_2}$ as before, such BRs are doubled.

\subsection{Event Rates}

In this section we combine the results from the Higgs boson cross-sections and
those from the BR analysis in order to perform a detailed study of typical event rates
for some Higgs signatures which are specific to the $B-L$ model.

Before all else, it is important to identify two different experimental
scenarios related to the LHC: we will generally refer to an
``early discovery scenario'' by considering an energy in the
hadronic Centre-of-Mass (CM) of $\sqrt{s}=7$ TeV and an integrated
luminosity of $\int{L}=1$~fb$^{-1}$ (according to the official schedule,
this is what is expected to be collected after the first couple of years of
LHC running) and to a ``full luminosity scenario'' by considering an
energy in the hadronic CM of $\sqrt{s}=14$ TeV and an
integrated luminosity of $\int L~=300$~fb$^{-1}$ (according to the official
schedule, this is what is expected to be realistically collected at the higher
energy stage).

As we shall see by combining the production cross-sections
and the decay BRs presented in the previous
subsections, the two different scenarios open different possibilities
for the detection of peculiar signatures of the model: in the
``early discovery scenario'' there is a clear possibility to
detect a light Higgs state yielding heavy neutrino pairs while the ``full luminosity
scenario'' affords the possibility of numerous
discovery mechanisms (in addition to the previous mechanism, for the heavy
Higgs state one also has decays of the latter  
into $Z'$ boson and light Higgs boson pairs).

\begin{figure}[!h]
  \subfloat[]{ 
  \label{gg-h1-nunu50}
  \includegraphics[angle=0,width=0.48\textwidth ]{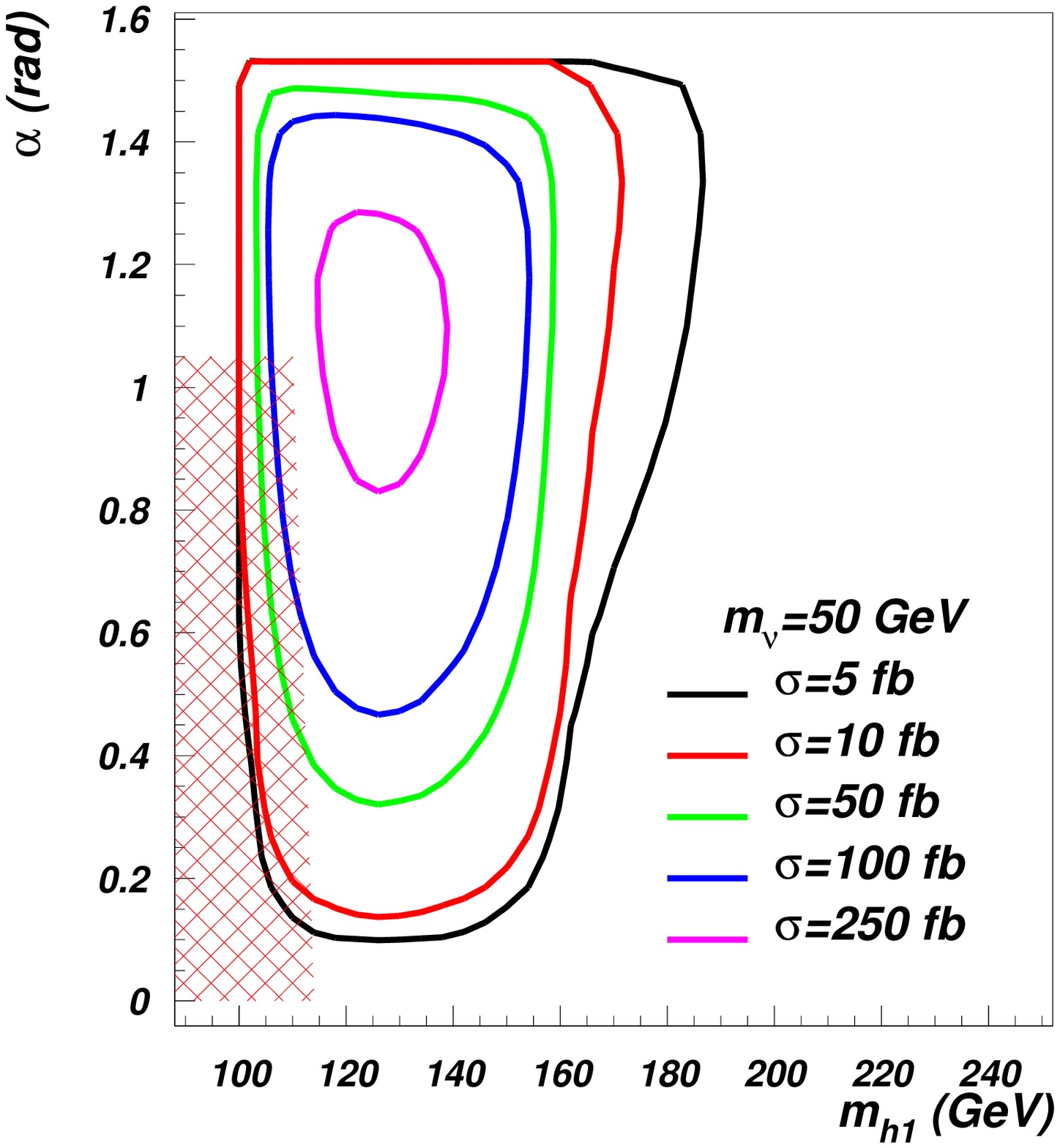}
}
  \subfloat[]{
  \label{gg-h1-nunu60}
  \includegraphics[angle=0,width=0.48\textwidth ]{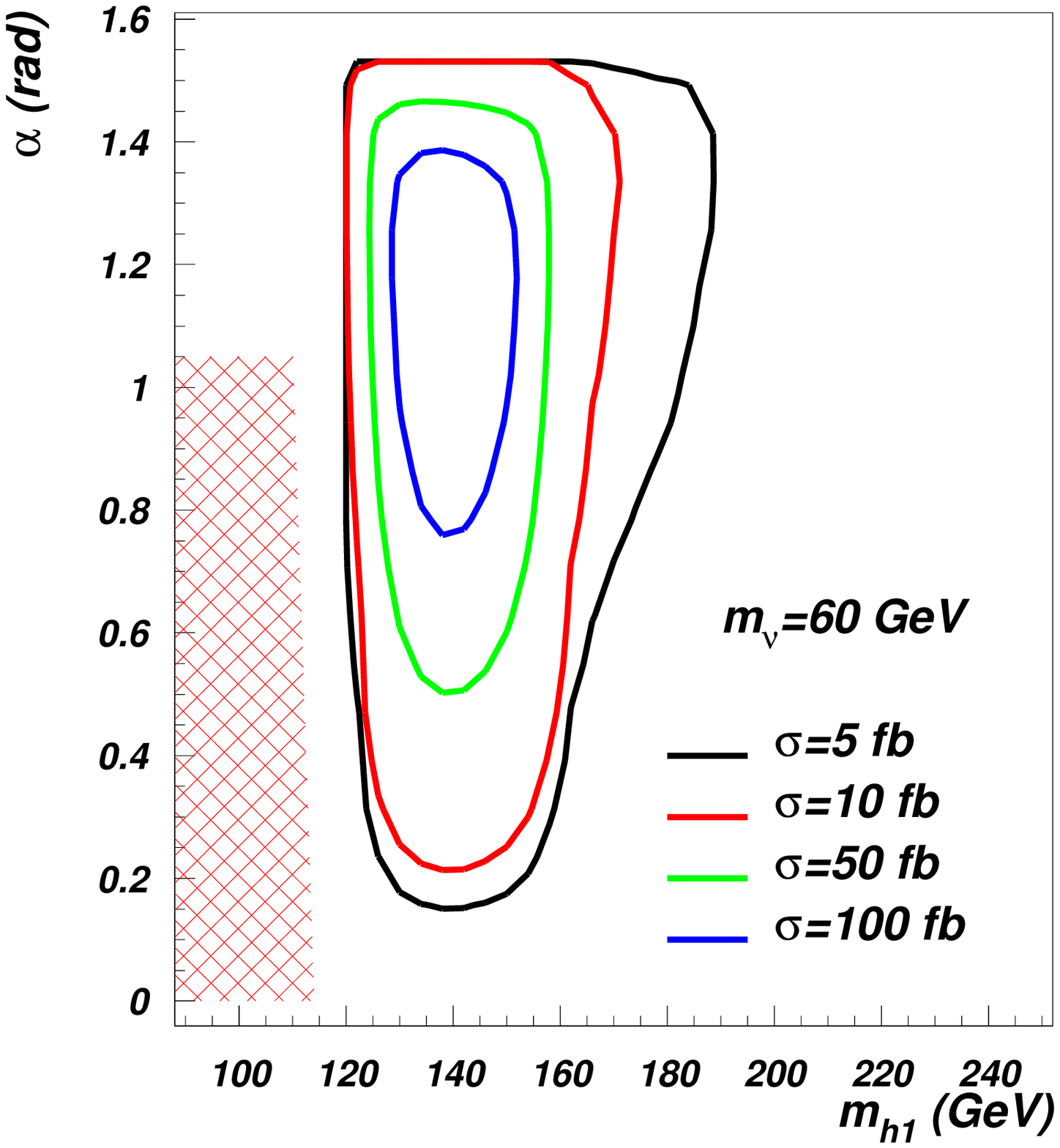}
}
  \vspace*{-0.5cm}
  \caption{\it Cross-section times BR contour 
    plot for the $B-L$ process $pp\rightarrow 
    h_1\rightarrow \nu_h \nu_h$ at the LHC with $\sqrt{s}=7$ TeV, plotted
    against $m_{h_1}$-$\alpha$, with $m_{\nu_h}=50$ GeV
    (\ref{gg-h1-nunu50}) and $m_{\nu_h}=60$ GeV
    (\ref{gg-h1-nunu60}). Several values of
    cross-section times BR have been considered: $\sigma=5$ fb (black line),
    $\sigma=10$ fb (red line), $\sigma=50$ fb (green line),
    $\sigma=100$ fb (blue line) and $\sigma=250$ fb (violet line). The
  red-shadowed region is excluded by the LEP experiments.}
  \label{ggnunu50-60}
\end{figure}

Firstly, we focus on the ``early discovery scenario'': in this
experimental configuration, the most important $B-L$ distinctive process is
represented by heavy neutrino pair production via a light Higgs boson, through the channel
$pp\rightarrow h_1 \rightarrow \nu_h\nu_h$.
In figure \ref{ggnunu50-60} we show the explicit results for
the $pp\rightarrow h_1\rightarrow \nu_h \nu_h$ process at the LHC with
$\sqrt{s}=7$ TeV, for $m_{\nu_h}=50$ GeV (figure (\ref{gg-h1-nunu50})) and
$m_{\nu_h}=60$ GeV (figure (\ref{gg-h1-nunu60})), obtained by combining the
light Higgs boson production cross-section via gluon-gluon fusion only 
(since it represents the main contribution) and the BR of the light Higgs boson to heavy neutrino pairs. The obtained rate is projected in the
$m_{h_1}$-$\alpha$ plane and several values of the
cross-section times BR have been considered: $\sigma=5$, $10$, $50$, $100$
and $250$ fb. The red-shadowed region takes into account the
exclusion limits established by the LEP experiments.

Even considering a low-luminosity scenario
(i.e., $\int L\simeq~1$~fb$^{-1}$), there is a
noticeable allowed parameter space for which the rate of such events is
considerably large: in the case of $m_{\nu_h}=50$ GeV, when the
integrated luminosity reaches $\int L=1$~fb$^{-1}$, we estimated a
collection of $\sim 10$ heavy neutrino pairs from 
the light Higgs boson production and decay for $100$ GeV$<m_{h_1}<170$ GeV and
$0.05\pi<\alpha<0.48\pi$, that scales up to $\sim 10^2$ events for
$110$ GeV$<m_{h_1}<155$ GeV and $0.16\pi<\alpha<0.46\pi$. In the case
of $m_{\nu_h}=60$ GeV, we estimated a
collection of $\sim 10$ heavy neutrino pairs from Higgs 
production for $120$ GeV$<m_{h_1}<170$ GeV and
$0.06\pi<\alpha<0.48\pi$, that scales up to $\sim 10^2$ events for
$125$ GeV$<m_{h_1}<150$ GeV and $0.25\pi<\alpha<0.44\pi$.

\begin{figure}[!h]
  \subfloat[]{ 
  \label{gg-h2-h1h1120}
  \includegraphics[angle=0,width=0.48\textwidth ]{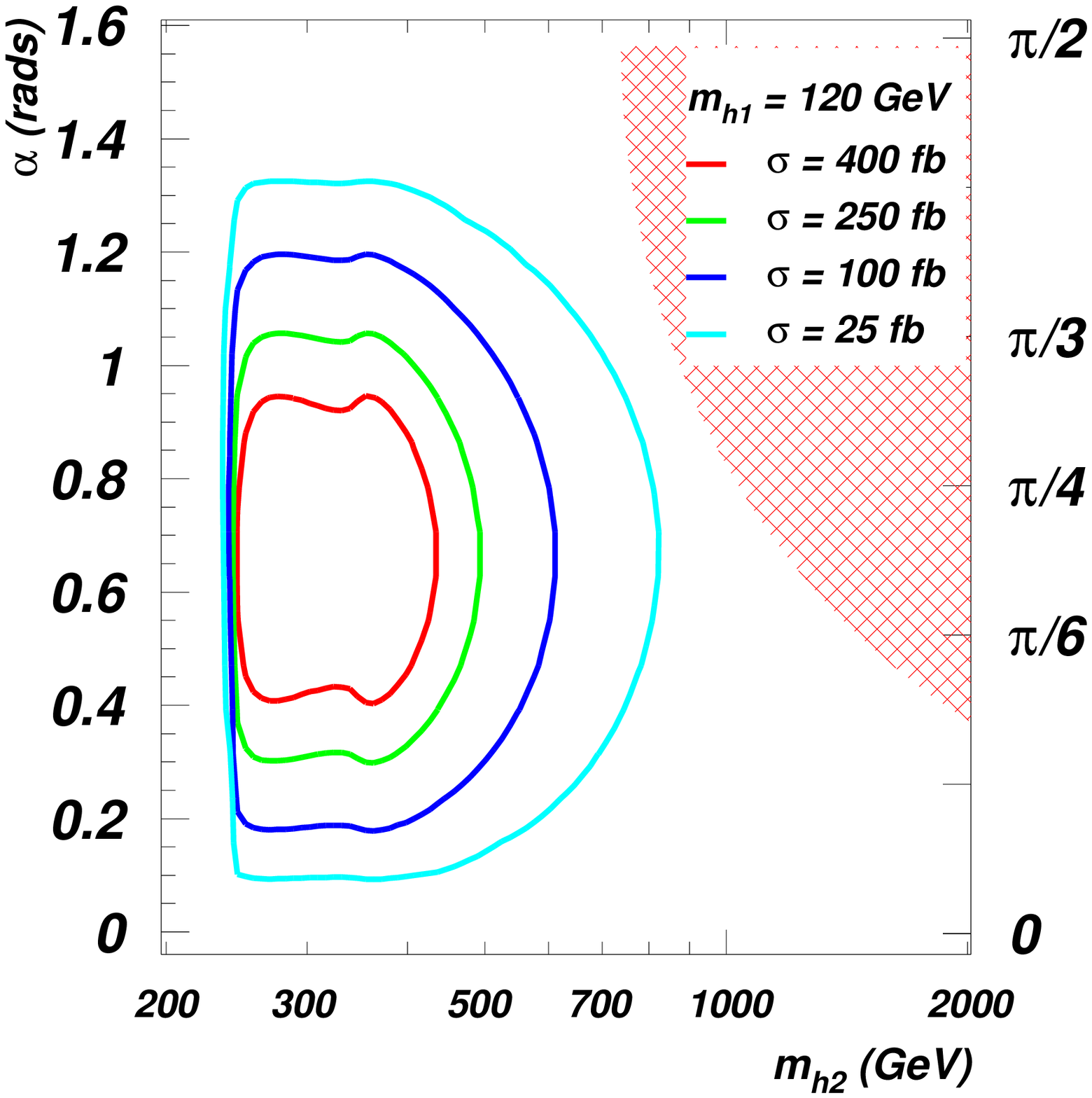}
}
  \subfloat[]{
  \label{gg-h2-h1h1240}
  \includegraphics[angle=0,width=0.48\textwidth ]{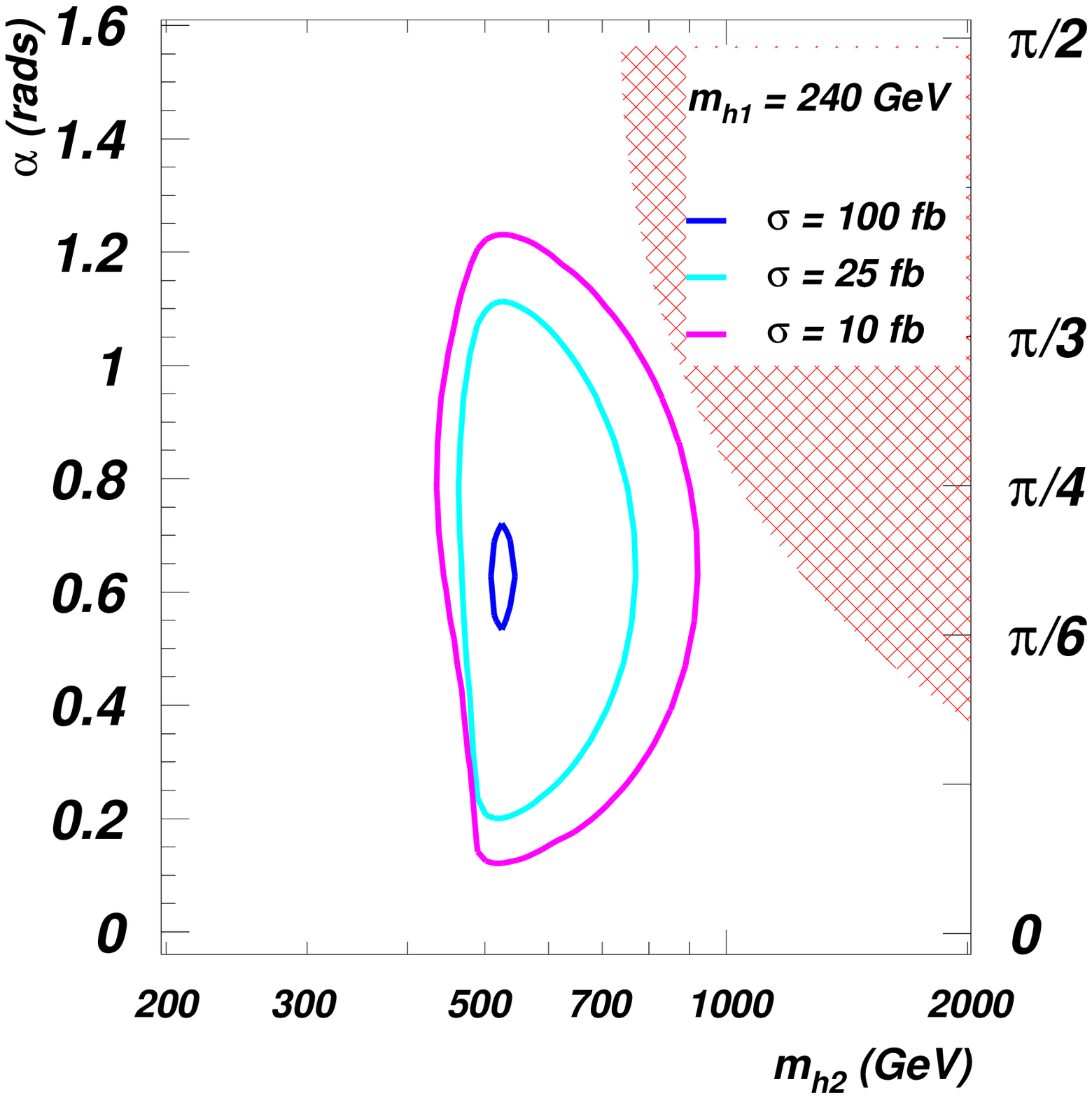}
}
  \vspace*{-0.5cm}
  \caption{\it Cross-section times BR contour 
    plot for the $B-L$ process $pp\rightarrow 
    h_2\rightarrow h_1 h_1$ at the LHC with $\sqrt{s}=14$ TeV, plotted
    against $m_{h_2}$-$\alpha$, with $m_{h_1}=120$
    GeV (\ref{gg-h2-h1h1120}) and $m_{h_1}=240$
    GeV (\ref{gg-h2-h1h1240}). Several values of 
    cross-section times BR have been considered: $\sigma=10$ fb (violet line),
    $\sigma=25$ fb (light-blue line), $\sigma=100$ fb (blue line),
    $\sigma=250$ fb (green line) and $\sigma=400$ fb (red line). The
  red-shadowed region is excluded by unitarity constraints.}
  \label{ggh1h1120-240}
\end{figure}

If we consider instead the ``full luminosity scenario'',
there are several important distinctive signatures: $pp\rightarrow h_2
\rightarrow h_1h_1$, $pp\rightarrow h_2 \rightarrow Z'Z'$ and
$pp\rightarrow h_2 \rightarrow \nu_h\nu_h$.
In figure \ref{ggh1h1120-240} we show the results for
light Higgs boson pair production from heavy Higgs boson decays at the LHC with
$\sqrt{s}=14$ TeV for $m_{h_1}=120$ GeV (figure (\ref{gg-h2-h1h1120})) and
$m_{h_1}=240$ GeV (figure (\ref{gg-h2-h1h1240})). Again, if we project
the rates on the bi-dimensional $m_{h_2}$-$\alpha$ plane, we can
select the contours that relate the cross-section times BR to some peculiar values.

Considering an integrated luminosity of $300$~fb$^{-1}$, we can
relate $\sigma=25(250)$ fb to $7500(75000)$ events, hence for both
choices of the light Higgs mass the $\alpha$-$m_{h_2}$ parameter space
offers an abundant portion in which the event rate is noticeable for
light Higgs boson pair production from heavy Higgs boson decays: when $m_{h_1}=120$ GeV the process is
accessible almost over the entire parameter space, with a cross-section
peak of $400$ fb in the
$240$ GeV$<m_{h_2}<400$ GeV and $0.13\pi<\alpha<0.30\pi$ intervals, while in the
$m_{h_1}=240$ GeV case the significant parameter space is still
large, even if slightly decreased, with a cross-section peak of
$25$ fb in the $480$ GeV$<m_{h_2}<800$ GeV and $0.06\pi<\alpha<0.32\pi$
region.

\begin{figure}[!h]
  \subfloat[]{ 
  \label{gg-h2-zpzp210}
  \includegraphics[angle=0,width=0.48\textwidth ]{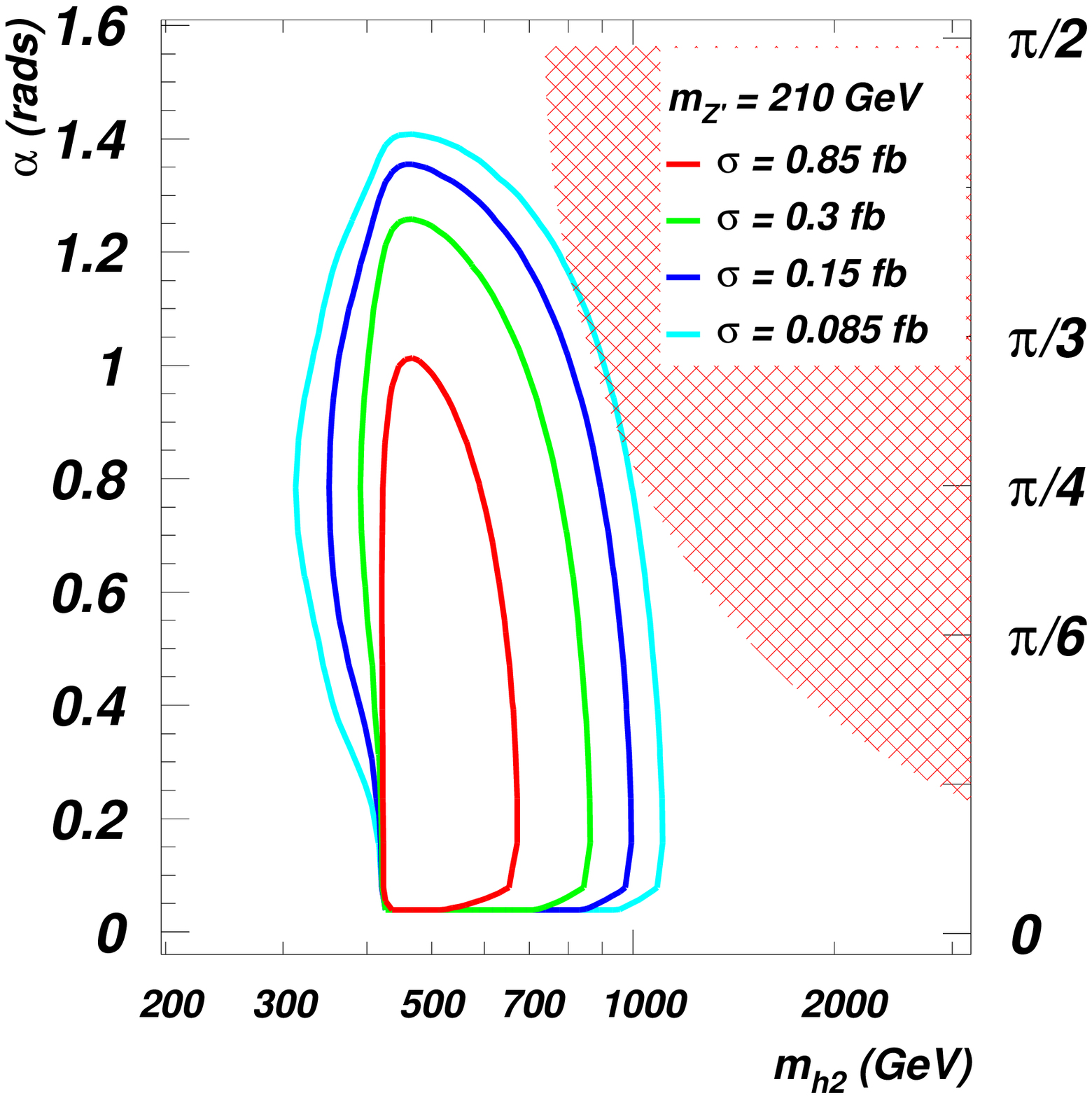}
}
  \subfloat[]{
  \label{gg-h2-zpzp280}
  \includegraphics[angle=0,width=0.48\textwidth ]{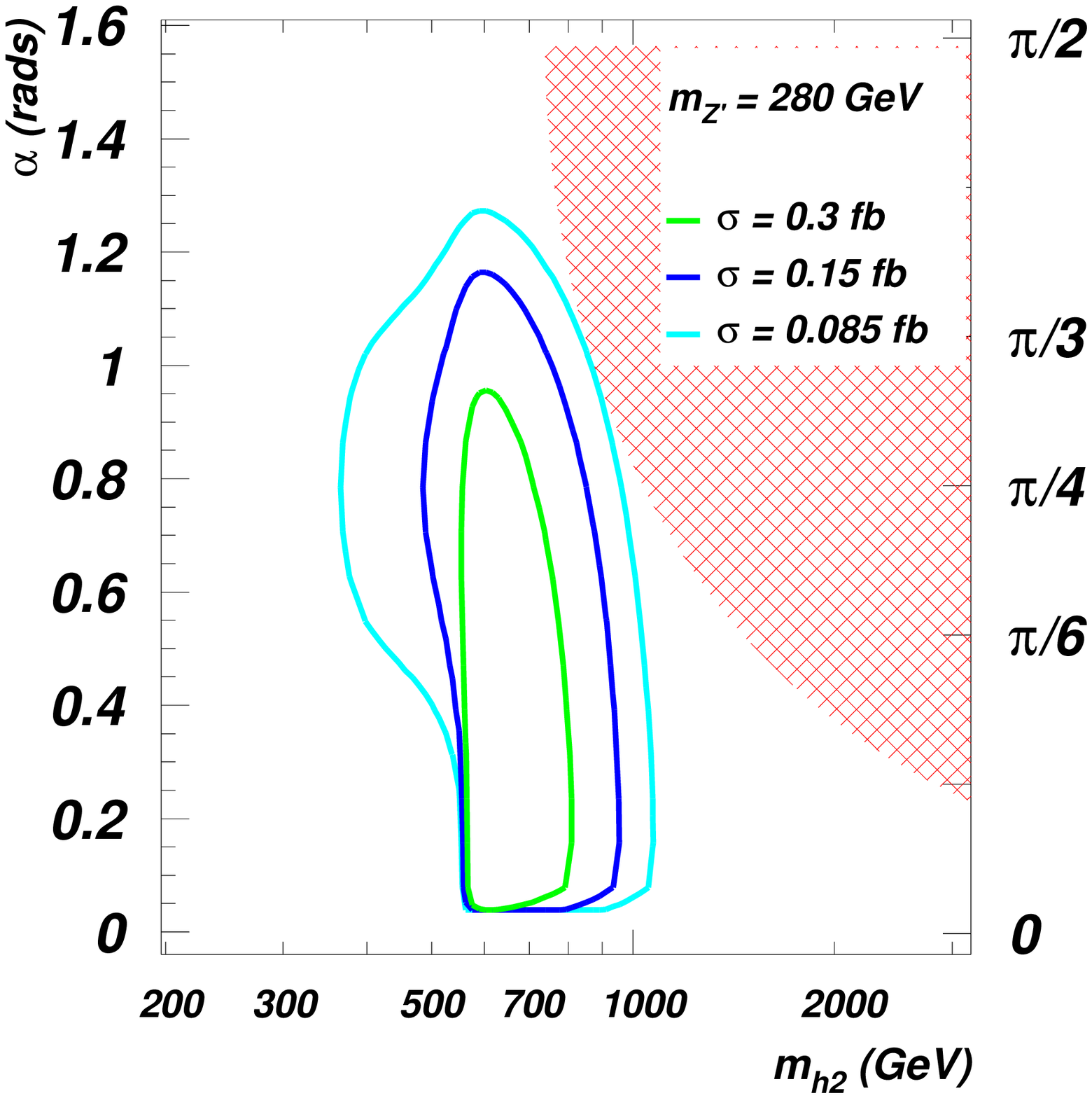}
}
  \vspace*{-0.5cm}
  \caption{\it Cross-section times BR contour 
    plot for the $B-L$ process $pp\rightarrow 
    h_2\rightarrow Z' Z'$ at the LHC with $\sqrt{s}=14$ TeV, plotted
    against $m_{h_2}$-$\alpha$, with $m_{Z'}=210$
    GeV (\ref{gg-h2-zpzp210}) and $m_{Z'}=280$
    GeV (\ref{gg-h2-zpzp280}). Several values of
    cross-section times BR have been considered: $\sigma=0.085$ fb (light-blue line),
    $\sigma=0.15$ fb (blue line), $\sigma=0.3$ fb (green line),
    $\sigma=0.85$ fb (red line). The
  red-shadowed region is excluded by unitarity constraints.}
  \label{ggzpzp210-280}
\end{figure}

In figure \ref{ggzpzp210-280} we show the results for
$Z'$ boson pair production from heavy Higgs boson decays at the LHC with
$\sqrt{s}=14$ TeV for $m_{Z'}=210$ GeV (figure (\ref{gg-h2-zpzp210})) and
$m_{Z'}=280$ GeV (figure (\ref{gg-h2-zpzp280})). Again, if we project
the rates on the bi-dimensional $m_{h_2}$-$\alpha$ plane, we can
select the contours that relate the cross-section times BR to some peculiar values.
Here, we have that $\sigma=0.085(0.85)$ fb corresponds to $25(250)$ events, hence for both
choices of $Z'$ mass the $\alpha$-$m_{h_2}$ parameter space
offers an abundant portion in which the event rate could be interesting for
$Z'$ boson pair production from heavy Higgs boson decays: for $m_{Z'}=210$ GeV the process 
has a peak of $0.85$ fb in the $420$
GeV$<m_{h_2}<650$ GeV and $0.03\pi<\alpha<0.25\pi$ region, while if
$m_{Z'}=280$ GeV a
noticeable parameter space is still potentially accessible with a rate
peak of $0.3$ fb ($100$ events) in the $560$ GeV$<m_{h_2}<800$ GeV and
$0.03\pi<\alpha<0.19\pi$ region.

\begin{figure}[!h]
  \subfloat[]{ 
  \label{gg-h2-nunu150}
  \includegraphics[angle=0,width=0.48\textwidth ]{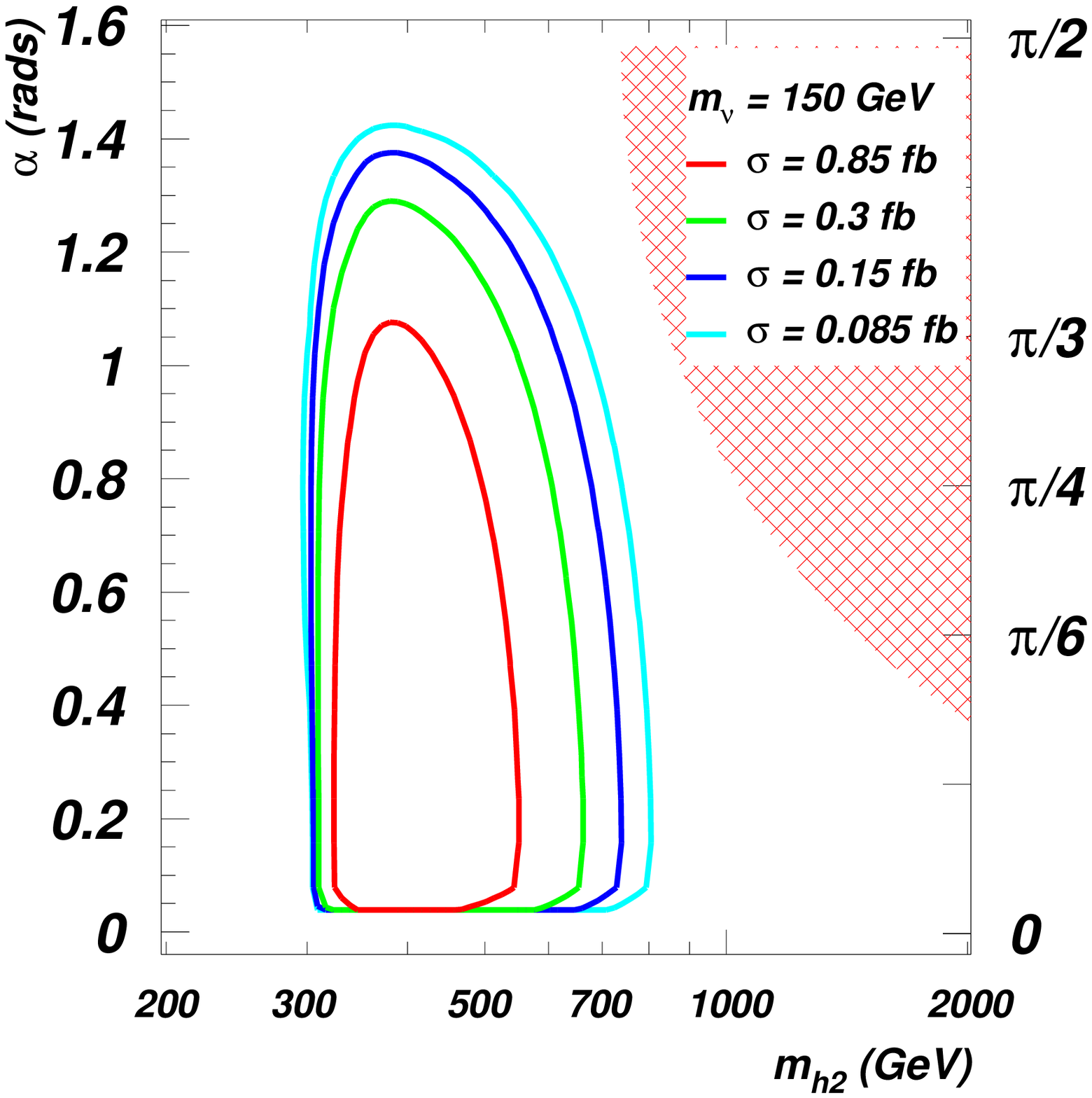}
}
  \subfloat[]{
  \label{gg-h2-nunu200}
  \includegraphics[angle=0,width=0.48\textwidth ]{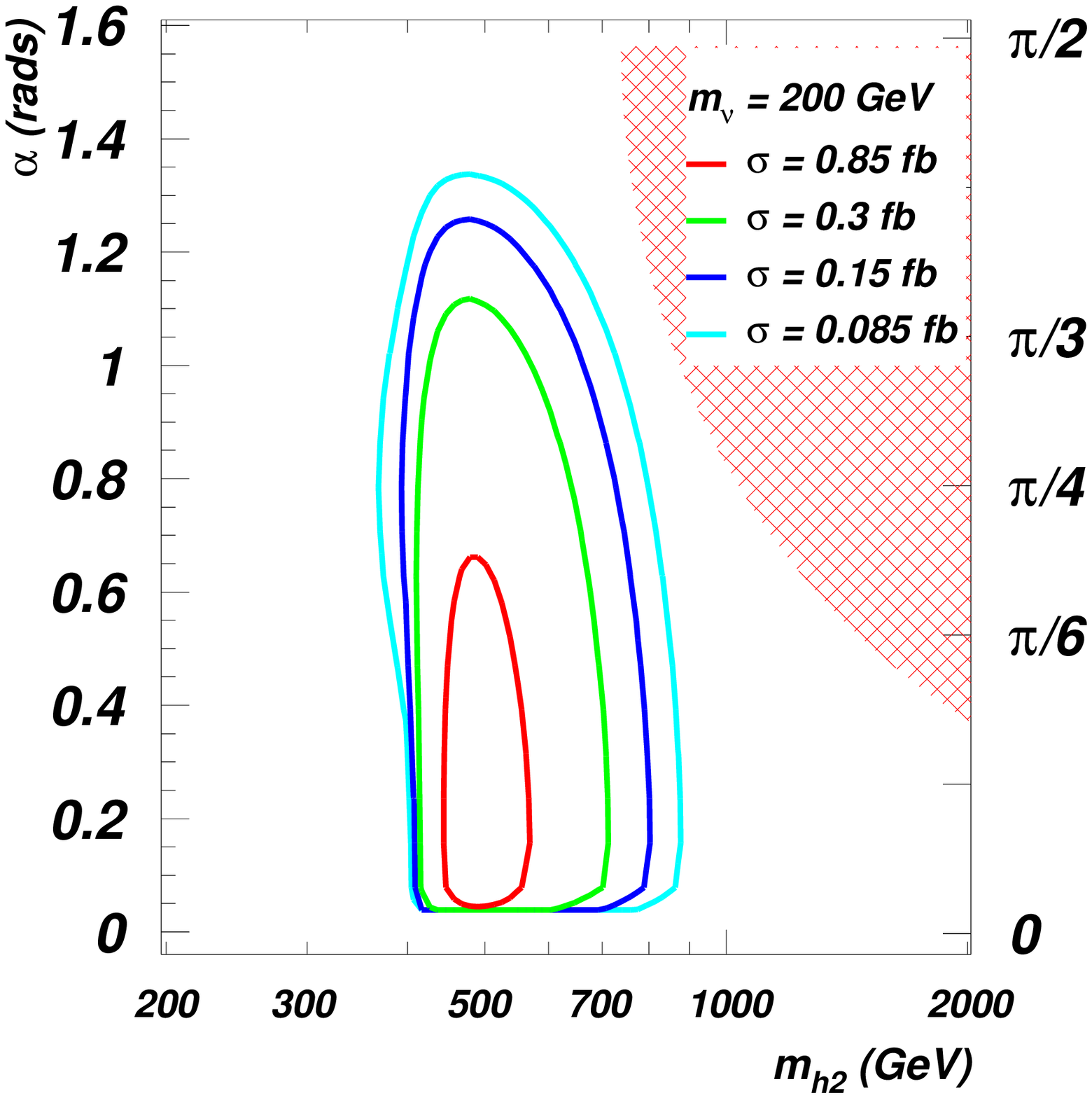}
}
  \vspace*{-0.5cm}
  \caption{\it Cross-section times BR contour 
    plot for the $B-L$ process $pp\rightarrow 
    h_2\rightarrow \nu_{h}\nu_{h}$ at the LHC with $\sqrt{s}=14$ TeV, plotted
    against $m_{h_2}$-$\alpha$, with $m_{\nu_{h}}=150$
    GeV (\ref{gg-h2-nunu150}) and $m_{\nu_{h}}=200$
    GeV (\ref{gg-h2-nunu200}). Several values of
    cross-section times BR have been considered: $\sigma=0.085$ fb (light-blue line),
    $\sigma=0.15$ fb (blue line), $\sigma=0.3$ fb (green line),
    $\sigma=0.85$ fb (red line). The
  red-shadowed region is excluded by unitarity constraints.}
  \label{ggnunu150-200}
\end{figure}

In analogy with the previous two cases, in figure \ref{ggnunu150-200} we
show the results for heavy neutrino pair production at the LHC with
$\sqrt{s}=14$ TeV plus $m_{\nu_h}=150$ GeV (figure (\ref{gg-h2-nunu150})) and
$m_{\nu_h}=200$ GeV (figure (\ref{gg-h2-nunu200})). The usual
contour plot displays a sizable event rate in the $\alpha$-$m_{h_2}$
parameter space for both choices of the $\nu_h$
mass. For example, when $m_{\nu_h}=150$ GeV we find a cross-section times BR 
peak of $0.85$ fb ($\sim 250$ events) in the $320$ GeV$<m_{h_2}<520$
GeV and $0.03\pi<\alpha<0.33\pi$ region, while if $m_{\nu_h}=200$ GeV we find
a peak of $0.85$ fb in the $450$ GeV$<m_{h_2}<550$ GeV
and $0.03\pi<\alpha<0.21\pi$ region.

\section{Conclusions}
\label{Sec:Conclusions}
In summary, we have studied in detail the Higgs sector of the minimal $B-L$ model at both
the foreseen energy stages of the LHC (and corresponding luminosities). While virtually all
relevant production and decay processes of the two Higgs states of the model have been investigated,
we have eventually paid particular attention 
to those that are peculiar to the described $B-L$ scenario. The phenomenological analysis
has been carried out in presence of all available theoretical and experimental constraints and by exploiting
numerical programs at the parton level. While many Higgs signatures already existing in the $SM$ could
be replicated in the case of its $B-L$ version, in either of the two Higgs states of the latter (depending 
on their mixing), it is more important to notice that several novel Higgs processes could act
as hallmarks of the minimal $B-L$ model. These include Higgs production via gluon-gluon fusion, 
in either the light or heavy Higgs state, the former produced at the lower energy stage of the CERN collider
and decaying in two heavy neutrinos and the latter produced at the higher energy stage of such a machine and
decaying not only in heavy neutrino pairs but also in $Z'$ and light Higgs ones. For each of these signatures
we have in fact found parameter space regions where the event rates are sizable and potentially amenable to
discovery. While, clearly, detailed signal-to-background analyses will have to either confirm or disprove
the possibility of the latter, our results have laid the basis for the phenomenological exploitation of
the Higgs sector of the minimal $B-L$ model at the LHC.


\section*{ACKNOWLEDGEMENTS}
\label{Sec:acknowledgements}

We would like to thank D.A.~Ross and C.H.~Shepherd-Themistocleous for most useful comments and discussions plus A.~Belyaev for the same and for his help in implementing the $gg$-fusion process and the appropriate Higgs boson widths  in CalcHEP as well as for suggesting the layout of figures~\ref{ggnunu50-60}-\ref{ggnunu150-200}. The work of all of us is supported in part by the NExT Institute.

\bibliography{biblio}

\begin{thebibliography}{10}%
\makeatletter
\providecommand \@ifxundefined [1]{%
 \ifx #1\undefined \expandafter \@firstoftwo
 \else \expandafter \@secondoftwo
\fi
}%
\providecommand \@ifnum [1]{%
 \ifnum #1\expandafter \@firstoftwo
 \else \expandafter \@secondoftwo
\fi
}%
\providecommand \enquote [1]{``#1''}%
\providecommand \bibnamefont  [1]{#1}%
\providecommand \bibfnamefont [1]{#1}%
\providecommand \citenamefont [1]{#1}%
\providecommand\href[0]{\@sanitize\@href}%
\providecommand\@href[1]{\endgroup\@@startlink{#1}\endgroup\@@href}%
\providecommand\@@href[1]{#1\@@endlink}%
\providecommand \@sanitize [0]{\begingroup\catcode`\&12\catcode`\#12\relax}%
\@ifxundefined \pdfoutput {\@firstoftwo}{%
 \@ifnum{\z@=\pdfoutput}{\@firstoftwo}{\@secondoftwo}%
}{%
 \providecommand\@@startlink[1]{\leavevmode\special{html:<a href="#1">}}%
 \providecommand\@@endlink[0]{\special{html:</a>}}%
}{%
 \providecommand\@@startlink[1]{%
  \leavevmode
  \pdfstartlink
   attr{/Border[0 0 1 ]/H/I/C[0 1 1]}%
   user{/Subtype/Link/A<</Type/Action/S/URI/URI(#1)>>}%
  \relax
 }%
 \providecommand\@@endlink[0]{\pdfendlink}%
}%
\providecommand \url  [0]{\begingroup\@sanitize \@url }%
\providecommand \@url [1]{\endgroup\@href {#1}{\urlprefix}}%
\providecommand \urlprefix [0]{URL }%
\providecommand \Eprint[0]{\href }%
\@ifxundefined \urlstyle {%
  \providecommand \doi [1]{doi:\discretionary{}{}{}#1}%
}{%
  \providecommand \doi [0]{doi:\discretionary{}{}{}\begingroup
  \urlstyle{rm}\Url }%
}%
\providecommand \doibase [0]{http://dx.doi.org/}%
\providecommand \Doi[1]{\href{\doibase#1}}%
\providecommand \bibAnnote [3]{%
  \BibitemShut{#1}%
  \begin{quotation}\noindent
    \textsc{Key:}\ #2\\\textsc{Annotation:}\ #3%
  \end{quotation}%
}%
\providecommand \bibAnnoteFile [2]{%
  \IfFileExists{#2}{\bibAnnote {#1} {#2} {\input{#2}}}{}%
}%
\providecommand \typeout [0]{\immediate \write \m@ne }%
\providecommand \selectlanguage [0]{\@gobble}%
\providecommand \bibinfo [0]{\@secondoftwo}%
\providecommand \bibfield [0]{\@secondoftwo}%
\providecommand \translation [1]{[#1]}%
\providecommand \BibitemOpen[0]{}%
\providecommand \bibitemStop [0]{}%
\providecommand \bibitemNoStop [0]{.\EOS\space}%
\providecommand \EOS [0]{\spacefactor3000\relax}%
\providecommand \BibitemShut [1]{\csname bibitem#1\endcsname}%
\bibitem{Djouadi:2005gi}%
  \BibitemOpen
  \bibfield{author}{%
  \bibinfo {author} {\bibfnamefont{A.}~\bibnamefont{Djouadi}},\ }%
  \bibfield{journal}{%
  \Doi{10.1016/j.physrep.2007.10.004}{\bibinfo {journal} {Phys. Rept.}}\ }%
  \textbf{\bibinfo {volume} {457}},\ \bibinfo {pages} {1} (\bibinfo {year}
  {2008}),\ \Eprint{http://arxiv.org/abs/hep-ph/0503172}{arXiv:hep-ph/0503172}%
  \bibAnnoteFile{NoStop}{Djouadi:2005gi}%
\bibitem{Jenkins:1987ue}%
  \BibitemOpen
  \bibfield{author}{%
  \bibinfo {author} {\bibfnamefont{E.~E.}\ \bibnamefont{Jenkins}},\ }%
  \bibfield{journal}{%
  \Doi{10.1016/0370-2693(87)91172-5}{\bibinfo {journal} {Phys. Lett.}}\ }%
  \textbf{\bibinfo {volume} {B192}},\ \bibinfo {pages} {219} (\bibinfo {year}
  {1987})%
  \bibAnnoteFile{NoStop}{Jenkins:1987ue}%
\bibitem{Buchmuller:1991ce}%
  \BibitemOpen
  \bibfield{author}{%
  \bibinfo {author} {\bibfnamefont{W.}~\bibnamefont{Buchmuller}}, \bibinfo
  {author} {\bibfnamefont{C.}~\bibnamefont{Greub}},\ and\ \bibinfo {author}
  {\bibfnamefont{P.}~\bibnamefont{Minkowski}},\ }%
  \bibfield{journal}{%
  \Doi{10.1016/0370-2693(91)90952-M}{\bibinfo {journal} {Phys. Lett.}}\ }%
  \textbf{\bibinfo {volume} {B267}},\ \bibinfo {pages} {395} (\bibinfo {year}
  {1991})%
  \bibAnnoteFile{NoStop}{Buchmuller:1991ce}%
\bibitem{Khalil:2006yi}%
  \BibitemOpen
  \bibfield{author}{%
  \bibinfo {author} {\bibfnamefont{S.}~\bibnamefont{Khalil}},\ }%
  \bibfield{journal}{%
  \Doi{10.1088/0954-3899/35/5/055001}{\bibinfo {journal} {J. Phys.}}\ }%
  \textbf{\bibinfo {volume} {G35}},\ \bibinfo {pages} {055001} (\bibinfo {year}
  {2008}),\ \Eprint{http://arxiv.org/abs/hep-ph/0611205}{arXiv:hep-ph/0611205}%
  \bibAnnoteFile{NoStop}{Khalil:2006yi}%
\bibitem{Minkowski:1977sc}%
  \BibitemOpen
  \bibfield{author}{%
  \bibinfo {author} {\bibfnamefont{P.}~\bibnamefont{Minkowski}},\ }%
  \bibfield{journal}{%
  \Doi{10.1016/0370-2693(77)90435-X}{\bibinfo {journal} {Phys. Lett.}}\ }%
  \textbf{\bibinfo {volume} {B67}},\ \bibinfo {pages} {421} (\bibinfo {year}
  {1977})%
  \bibAnnoteFile{NoStop}{Minkowski:1977sc}%
\bibitem{VanNieuwenhuizen:1979hm}%
  \BibitemOpen
  \bibfield{author}{%
  \bibinfo {author} {\bibfnamefont{P.}~\bibnamefont{Van~Nieuwenhuizen}}\ and\
  \bibinfo {author} {\bibfnamefont{D.~Z.}\ \bibnamefont{Freedman}},\ \bibinfo
  {pages} {341}}%
   (\bibinfo {year} {1979}),\ \bibinfo {note} {amsterdam, Netherlands:
  North-Holland}%
  \bibAnnoteFile{NoStop}{VanNieuwenhuizen:1979hm}%
\bibitem{Yanagida:1979as}%
  \BibitemOpen
  \bibfield{author}{%
  \bibinfo {author} {\bibfnamefont{T.}~\bibnamefont{Yanagida}}}%
   (\bibinfo {year} {1979}),\ \bibinfo {note} {in Proceedings of the Workshop
  on the Baryon Number of the Universe and Unified Theories, Tsukuba, Japan,
  13-14 Feb}%
  \bibAnnoteFile{NoStop}{Yanagida:1979as}%
\bibitem{GellMann:1980vs}%
  \BibitemOpen
  \bibfield{author}{%
  \bibinfo {author} {\bibfnamefont{M.}~\bibnamefont{Gell-Mann}}, \bibinfo
  {author} {\bibfnamefont{P.}~\bibnamefont{Ramond}},\ and\ \bibinfo {author}
  {\bibfnamefont{R.}~\bibnamefont{Slansky}}\ }%
  \bibinfo {note} {print-80-0576 (CERN)}%
  \bibAnnoteFile{NoStop}{GellMann:1980vs}%
\bibitem{S.L.Glashow}%
  \BibitemOpen
  \bibinfo {note} {S.L. Glashow, in \emph{Quarks and Leptons}, eds. M.L\`evy et
  al. (Plenum, New York $1980$), p.~$707$}%
  \bibAnnoteFile{NoStop}{S.L.Glashow}%
\bibitem{Mohapatra:1979ia}%
  \BibitemOpen
  \bibfield{author}{%
  \bibinfo {author} {\bibfnamefont{R.~N.}\ \bibnamefont{Mohapatra}}\ and\
  \bibinfo {author} {\bibfnamefont{G.}~\bibnamefont{Senjanovic}},\ }%
  \bibfield{journal}{%
  \Doi{10.1103/PhysRevLett.44.912}{\bibinfo {journal} {Phys. Rev. Lett.}}\ }%
  \textbf{\bibinfo {volume} {44}},\ \bibinfo {pages} {912} (\bibinfo {year}
  {1980})%
  \bibAnnoteFile{NoStop}{Mohapatra:1979ia}%
\bibitem{Emam:2007dy}%
  \BibitemOpen
  \bibfield{author}{%
  \bibinfo {author} {\bibfnamefont{W.}~\bibnamefont{Emam}}\ and\ \bibinfo
  {author} {\bibfnamefont{S.}~\bibnamefont{Khalil}},\ }%
  \bibfield{journal}{%
  \Doi{10.1140/epjc/s10052-007-0411-7}{\bibinfo {journal} {Eur. Phys. J.}}\ }%
  \textbf{\bibinfo {volume} {C52}},\ \bibinfo {pages} {625} (\bibinfo {year}
  {2007}),\ \Eprint{http://arxiv.org/abs/0704.1395}{arXiv:0704.1395 [hep-ph]}%
  \bibAnnoteFile{NoStop}{Emam:2007dy}%
\bibitem{Basso:2008iv}%
  \BibitemOpen
  \bibfield{author}{%
  \bibinfo {author} {\bibfnamefont{L.}~\bibnamefont{Basso}}, \bibinfo {author}
  {\bibfnamefont{A.}~\bibnamefont{Belyaev}}, \bibinfo {author}
  {\bibfnamefont{S.}~\bibnamefont{Moretti}},\ and\ \bibinfo {author}
  {\bibfnamefont{C.~H.}\ \bibnamefont{Shepherd-Themistocleous}},\ }%
  \bibfield{journal}{%
  \Doi{10.1103/PhysRevD.80.055030}{\bibinfo {journal} {Phys. Rev.}}\ }%
  \textbf{\bibinfo {volume} {D80}},\ \bibinfo {pages} {055030} (\bibinfo {year}
  {2009}),\
  \Eprint{http://arxiv.org/abs/hep-ph/0812.4313}{arXiv:hep-ph/0812.4313}%
  \bibAnnoteFile{NoStop}{Basso:2008iv}%
\bibitem{Emam:2008zz}%
  \BibitemOpen
  \bibfield{author}{%
  \bibinfo {author} {\bibfnamefont{W.}~\bibnamefont{Emam}}\ and\ \bibinfo
  {author} {\bibfnamefont{P.}~\bibnamefont{Mine}},\ }%
  \bibfield{journal}{%
  \Doi{10.1088/0954-3899/36/12/129701}{\bibinfo {journal} {J. Phys.}}\ }%
  \textbf{\bibinfo {volume} {G36}},\ \bibinfo {pages} {129701} (\bibinfo {year}
  {2009})%
  \bibAnnoteFile{NoStop}{Emam:2008zz}%
\bibitem{Huitu:2008gf}%
  \BibitemOpen
  \bibfield{author}{%
  \bibinfo {author} {\bibfnamefont{K.}~\bibnamefont{Huitu}}, \bibinfo {author}
  {\bibfnamefont{S.}~\bibnamefont{Khalil}}, \bibinfo {author}
  {\bibfnamefont{H.}~\bibnamefont{Okada}},\ and\ \bibinfo {author}
  {\bibfnamefont{S.~K.}\ \bibnamefont{Rai}},\ }%
  \bibfield{journal}{%
  \Doi{10.1103/PhysRevLett.101.181802}{\bibinfo {journal} {Phys. Rev. Lett.}}\
  }%
  \textbf{\bibinfo {volume} {101}},\ \bibinfo {pages} {181802} (\bibinfo {year}
  {2008}),\ \Eprint{http://arxiv.org/abs/0803.2799}{arXiv:0803.2799 [hep-ph]}%
  \bibAnnoteFile{NoStop}{Huitu:2008gf}%
\bibitem{Basso:2009hf}%
  \BibitemOpen
  \bibfield{author}{%
  \bibinfo {author} {\bibfnamefont{L.}~\bibnamefont{Basso}}, \bibinfo {author}
  {\bibfnamefont{A.}~\bibnamefont{Belyaev}}, \bibinfo {author}
  {\bibfnamefont{S.}~\bibnamefont{Moretti}},\ and\ \bibinfo {author}
  {\bibfnamefont{G.~M.}\ \bibnamefont{Pruna}},\ }%
  \bibfield{journal}{%
  \Doi{10.1088/1126-6708/2009/10/006}{\bibinfo {journal} {JHEP}}\ }%
  \textbf{\bibinfo {volume} {10}},\ \bibinfo {pages} {006} (\bibinfo {year}
  {2009}),\ \Eprint{http://arxiv.org/abs/0903.4777}{arXiv:0903.4777 [hep-ph]}%
  \bibAnnoteFile{NoStop}{Basso:2009hf}%
\bibitem{Basso:2010pe}%
  \BibitemOpen
  \bibfield{author}{%
  \bibinfo {author} {\bibfnamefont{L.}~\bibnamefont{Basso}}, \bibinfo {author}
  {\bibfnamefont{A.}~\bibnamefont{Belyaev}}, \bibinfo {author}
  {\bibfnamefont{S.}~\bibnamefont{Moretti}}, \bibinfo {author}
  {\bibfnamefont{G.~M.}\ \bibnamefont{Pruna}},\ and\ \bibinfo {author}
  {\bibfnamefont{C.~H.}\ \bibnamefont{Shepherd-Themistocleous}}}%
   (\bibinfo {year} {2010}),\
  \Eprint{http://arxiv.org/abs/1002.3586}{arXiv:1002.3586 [hep-ph]}%
  \bibAnnoteFile{NoStop}{Basso:2010pe}%
\bibitem{BahatTreidel:2006kx}%
  \BibitemOpen
  \bibfield{author}{%
  \bibinfo {author} {\bibfnamefont{O.}~\bibnamefont{Bahat-Treidel}}, \bibinfo
  {author} {\bibfnamefont{Y.}~\bibnamefont{Grossman}},\ and\ \bibinfo {author}
  {\bibfnamefont{Y.}~\bibnamefont{Rozen}},\ }%
  \bibfield{journal}{%
  \Doi{10.1088/1126-6708/2007/05/022}{\bibinfo {journal} {JHEP}}\ }%
  \textbf{\bibinfo {volume} {05}},\ \bibinfo {pages} {022} (\bibinfo {year}
  {2007}),\ \Eprint{http://arxiv.org/abs/hep-ph/0611162}{arXiv:hep-ph/0611162}%
  \bibAnnoteFile{NoStop}{BahatTreidel:2006kx}%
\bibitem{O'Connell:2006wi}%
  \BibitemOpen
  \bibfield{author}{%
  \bibinfo {author} {\bibfnamefont{D.}~\bibnamefont{O'Connell}}, \bibinfo
  {author} {\bibfnamefont{M.~J.}\ \bibnamefont{Ramsey-Musolf}},\ and\ \bibinfo
  {author} {\bibfnamefont{M.~B.}\ \bibnamefont{Wise}},\ }%
  \bibfield{journal}{%
  \Doi{10.1103/PhysRevD.75.037701}{\bibinfo {journal} {Phys. Rev.}}\ }%
  \textbf{\bibinfo {volume} {D75}},\ \bibinfo {pages} {037701} (\bibinfo {year}
  {2007}),\ \Eprint{http://arxiv.org/abs/hep-ph/0611014}{arXiv:hep-ph/0611014}%
  \bibAnnoteFile{NoStop}{O'Connell:2006wi}%
\bibitem{Barger:2007im}%
  \BibitemOpen
  \bibfield{author}{%
  \bibinfo {author} {\bibfnamefont{V.}~\bibnamefont{Barger}}, \bibinfo {author}
  {\bibfnamefont{P.}~\bibnamefont{Langacker}}, \bibinfo {author}
  {\bibfnamefont{M.}~\bibnamefont{McCaskey}}, \bibinfo {author}
  {\bibfnamefont{M.~J.}\ \bibnamefont{Ramsey-Musolf}},\ and\ \bibinfo {author}
  {\bibfnamefont{G.}~\bibnamefont{Shaughnessy}},\ }%
  \bibfield{journal}{%
  \Doi{10.1103/PhysRevD.77.035005}{\bibinfo {journal} {Phys. Rev.}}\ }%
  \textbf{\bibinfo {volume} {D77}},\ \bibinfo {pages} {035005} (\bibinfo {year}
  {2008}),\ \Eprint{http://arxiv.org/abs/0706.4311}{arXiv:0706.4311 [hep-ph]}%
  \bibAnnoteFile{NoStop}{Barger:2007im}%
\bibitem{Bhattacharyya:2007pb}%
  \BibitemOpen
  \bibfield{author}{%
  \bibinfo {author} {\bibfnamefont{G.}~\bibnamefont{Bhattacharyya}}, \bibinfo
  {author} {\bibfnamefont{G.~C.}\ \bibnamefont{Branco}},\ and\ \bibinfo
  {author} {\bibfnamefont{S.}~\bibnamefont{Nandi}},\ }%
  \bibfield{journal}{%
  \Doi{10.1103/PhysRevD.77.117701}{\bibinfo {journal} {Phys. Rev.}}\ }%
  \textbf{\bibinfo {volume} {D77}},\ \bibinfo {pages} {117701} (\bibinfo {year}
  {2008}),\ \Eprint{http://arxiv.org/abs/0712.2693}{arXiv:0712.2693 [hep-ph]}%
  \bibAnnoteFile{NoStop}{Bhattacharyya:2007pb}%
\bibitem{Profumo:2007wc}%
  \BibitemOpen
  \bibfield{author}{%
  \bibinfo {author} {\bibfnamefont{S.}~\bibnamefont{Profumo}}, \bibinfo
  {author} {\bibfnamefont{M.~J.}\ \bibnamefont{Ramsey-Musolf}},\ and\ \bibinfo
  {author} {\bibfnamefont{G.}~\bibnamefont{Shaughnessy}},\ }%
  \bibfield{journal}{%
  \Doi{10.1088/1126-6708/2007/08/010}{\bibinfo {journal} {JHEP}}\ }%
  \textbf{\bibinfo {volume} {08}},\ \bibinfo {pages} {010} (\bibinfo {year}
  {2007}),\ \Eprint{http://arxiv.org/abs/0705.2425}{arXiv:0705.2425 [hep-ph]}%
  \bibAnnoteFile{NoStop}{Profumo:2007wc}%
\bibitem{Barger:2008jx}%
  \BibitemOpen
  \bibfield{author}{%
  \bibinfo {author} {\bibfnamefont{V.}~\bibnamefont{Barger}}, \bibinfo {author}
  {\bibfnamefont{P.}~\bibnamefont{Langacker}}, \bibinfo {author}
  {\bibfnamefont{M.}~\bibnamefont{McCaskey}}, \bibinfo {author}
  {\bibfnamefont{M.}~\bibnamefont{Ramsey-Musolf}},\ and\ \bibinfo {author}
  {\bibfnamefont{G.}~\bibnamefont{Shaughnessy}},\ }%
  \bibfield{journal}{%
  \Doi{10.1103/PhysRevD.79.015018}{\bibinfo {journal} {Phys. Rev.}}\ }%
  \textbf{\bibinfo {volume} {D79}},\ \bibinfo {pages} {015018} (\bibinfo {year}
  {2009}),\ \Eprint{http://arxiv.org/abs/0811.0393}{arXiv:0811.0393 [hep-ph]}%
  \bibAnnoteFile{NoStop}{Barger:2008jx}%
\bibitem{Basso:2010jt}%
  \BibitemOpen
  \bibfield{author}{%
  \bibinfo {author} {\bibfnamefont{L.}~\bibnamefont{Basso}}, \bibinfo {author}
  {\bibfnamefont{A.}~\bibnamefont{Belyaev}}, \bibinfo {author}
  {\bibfnamefont{S.}~\bibnamefont{Moretti}},\ and\ \bibinfo {author}
  {\bibfnamefont{G.~M.}\ \bibnamefont{Pruna}},\ }%
  \bibfield{journal}{%
  \Doi{10.1103/PhysRevD.81.095018}{\bibinfo {journal} {Phys. Rev.}}\ }%
  \textbf{\bibinfo {volume} {D81}},\ \bibinfo {pages} {095018} (\bibinfo {year}
  {2010}),\ \Eprint{http://arxiv.org/abs/1002.1939}{arXiv:1002.1939 [hep-ph]}%
  \bibAnnoteFile{NoStop}{Basso:2010jt}%
\bibitem{Basso:2010jm}%
  \BibitemOpen
  \bibfield{author}{%
  \bibinfo {author} {\bibfnamefont{L.}~\bibnamefont{Basso}}, \bibinfo {author}
  {\bibfnamefont{S.}~\bibnamefont{Moretti}},\ and\ \bibinfo {author}
  {\bibfnamefont{G.~M.}\ \bibnamefont{Pruna}},\ }%
  \bibfield{journal}{%
  \bibinfo {journal} {Phys. Rev.}\ }%
  \textbf{\bibinfo {volume} {D82}},\ \bibinfo {pages} {055018} (\bibinfo {year}
  {2010}),\ \Eprint{http://arxiv.org/abs/1004.3039}{arXiv:1004.3039 [hep-ph]}%
  \bibAnnoteFile{NoStop}{Basso:2010jm}%
\bibitem{Basso:2010hk}%
  \BibitemOpen
  \bibfield{author}{%
  \bibinfo {author} {\bibfnamefont{L.}~\bibnamefont{Basso}}, \bibinfo {author}
  {\bibfnamefont{S.}~\bibnamefont{Moretti}},\ and\ \bibinfo {author}
  {\bibfnamefont{G.~M.}\ \bibnamefont{Pruna}}}%
   (\bibinfo {year} {2010}),\
  \Eprint{http://arxiv.org/abs/1009.4164}{arXiv:1009.4164 [hep-ph]}%
  \bibAnnoteFile{NoStop}{Basso:2010hk}%
\bibitem{Dawson:2009yx}%
  \BibitemOpen
  \bibfield{author}{%
  \bibinfo {author} {\bibfnamefont{S.}~\bibnamefont{Dawson}}\ and\ \bibinfo
  {author} {\bibfnamefont{W.}~\bibnamefont{Yan}},\ }%
  \bibfield{journal}{%
  \Doi{10.1103/PhysRevD.79.095002}{\bibinfo {journal} {Phys. Rev.}}\ }%
  \textbf{\bibinfo {volume} {D79}},\ \bibinfo {pages} {095002} (\bibinfo {year}
  {2009}),\ \Eprint{http://arxiv.org/abs/0904.2005}{arXiv:0904.2005 [hep-ph]}%
  \bibAnnoteFile{NoStop}{Dawson:2009yx}%
\bibitem{Cacciapaglia:2006pk}%
  \BibitemOpen
  \bibfield{author}{%
  \bibinfo {author} {\bibfnamefont{G.}~\bibnamefont{Cacciapaglia}}, \bibinfo
  {author} {\bibfnamefont{C.}~\bibnamefont{Csaki}}, \bibinfo {author}
  {\bibfnamefont{G.}~\bibnamefont{Marandella}},\ and\ \bibinfo {author}
  {\bibfnamefont{A.}~\bibnamefont{Strumia}},\ }%
  \bibfield{journal}{%
  \Doi{10.1103/PhysRevD.74.033011}{\bibinfo {journal} {Phys. Rev.}}\ }%
  \textbf{\bibinfo {volume} {D74}},\ \bibinfo {pages} {033011} (\bibinfo {year}
  {2006}),\ \Eprint{http://arxiv.org/abs/hep-ph/0604111}{arXiv:hep-ph/0604111}%
  \bibAnnoteFile{NoStop}{Cacciapaglia:2006pk}%
\bibitem{Anthony:2003ub}%
  \BibitemOpen
  \bibfield{author}{%
  \bibinfo {author} {\bibfnamefont{P.~L.}\ \bibnamefont{Anthony}} \emph{et~al.}
  (\bibinfo {collaboration} {SLAC E158}),\ }%
  \bibfield{journal}{%
  \Doi{10.1103/PhysRevLett.92.181602}{\bibinfo {journal} {Phys. Rev. Lett.}}\
  }%
  \textbf{\bibinfo {volume} {92}},\ \bibinfo {pages} {181602} (\bibinfo {year}
  {2004}),\ \Eprint{http://arxiv.org/abs/hep-ex/0312035}{arXiv:hep-ex/0312035}%
  \bibAnnoteFile{NoStop}{Anthony:2003ub}%
\bibitem{ew:2003ih}%
  \BibitemOpen
  \bibfield{author}{%
  \bibinfo {author} {\bibnamefont{{The ALEPH, DELPHI, L3, OPAL, SLD
  Collaborations, the LEP Electroweak Working Group, the SLD Electroweak and
  Heavy Flavour Groups}}} (\bibinfo {collaboration} {LEP})}%
   (\bibinfo {year} {2003}),\
  \Eprint{http://arxiv.org/abs/hep-ex/0312023}{arXiv:hep-ex/0312023}%
  \bibAnnoteFile{NoStop}{ew:2003ih}%
\bibitem{Azzi:2004rc}%
  \BibitemOpen
  \bibfield{author}{%
  \bibinfo {author} {\bibfnamefont{P.}~\bibnamefont{Azzi}} \emph{et~al.}
  (\bibinfo {collaboration} {CDF and D0 and Tevatron Electroweak Working
  Group})}%
   (\bibinfo {year} {2004}),\
  \Eprint{http://arxiv.org/abs/hep-ex/0404010}{arXiv:hep-ex/0404010}%
  \bibAnnoteFile{NoStop}{Azzi:2004rc}%
\bibitem{Woods:2004zr}%
  \BibitemOpen
  \bibfield{author}{%
  \bibinfo {author} {\bibfnamefont{M.}~\bibnamefont{Woods}} (\bibinfo
  {collaboration} {SLAC E158})}%
   (\bibinfo {year} {2004}),\
  \Eprint{http://arxiv.org/abs/hep-ex/0403010}{arXiv:hep-ex/0403010}%
  \bibAnnoteFile{NoStop}{Woods:2004zr}%
\bibitem{Z-Pole}%
  \BibitemOpen
  \bibfield{author}{%
  \bibinfo {author} {\bibnamefont{{The ALEPH, DELPHI, L3, OPAL, SLD
  Collaborations, the LEP Electroweak Working Group, the SLD Electroweak and
  Heavy Flavour Groups}}},\ }%
  \bibfield{journal}{%
  \bibinfo {journal} {Phys. Rept.}\ }%
  \textbf{\bibinfo {volume} {427}},\ \bibinfo {pages} {257} (\bibinfo {year}
  {2006}),\ \Eprint{http://arxiv.org/abs/hep-ex/0509008}{hep-ex/0509008}%
  \bibAnnoteFile{NoStop}{Z-Pole}%
\bibitem{Carena:2004xs}%
  \BibitemOpen
  \bibfield{author}{%
  \bibinfo {author} {\bibfnamefont{M.~S.}\ \bibnamefont{Carena}}, \bibinfo
  {author} {\bibfnamefont{A.}~\bibnamefont{Daleo}}, \bibinfo {author}
  {\bibfnamefont{B.~A.}\ \bibnamefont{Dobrescu}},\ and\ \bibinfo {author}
  {\bibfnamefont{T.~M.~P.}\ \bibnamefont{Tait}},\ }%
  \bibfield{journal}{%
  \Doi{10.1103/PhysRevD.70.093009}{\bibinfo {journal} {Phys. Rev.}}\ }%
  \textbf{\bibinfo {volume} {D70}},\ \bibinfo {pages} {093009} (\bibinfo {year}
  {2004}),\ \Eprint{http://arxiv.org/abs/hep-ph/0408098}{arXiv:hep-ph/0408098}%
  \bibAnnoteFile{NoStop}{Carena:2004xs}%
\bibitem{Aaltonen:2008vx}%
  \BibitemOpen
  \bibfield{author}{%
  \bibinfo {author} {\bibfnamefont{T.}~\bibnamefont{Aaltonen}} \emph{et~al.}
  (\bibinfo {collaboration} {CDF}),\ }%
  \bibfield{journal}{%
  \Doi{10.1103/PhysRevLett.102.031801}{\bibinfo {journal} {Phys. Rev. Lett.}}\
  }%
  \textbf{\bibinfo {volume} {102}},\ \bibinfo {pages} {031801} (\bibinfo {year}
  {2009}),\ \Eprint{http://arxiv.org/abs/0810.2059}{arXiv:0810.2059 [hep-ex]}%
  \bibAnnoteFile{NoStop}{Aaltonen:2008vx}%
\bibitem{Aaltonen:2008ah}%
  \BibitemOpen
  \bibfield{author}{%
  \bibinfo {author} {\bibfnamefont{T.}~\bibnamefont{Aaltonen}} \emph{et~al.}
  (\bibinfo {collaboration} {CDF}),\ }%
  \bibfield{journal}{%
  \Doi{10.1103/PhysRevLett.102.091805}{\bibinfo {journal} {Phys. Rev. Lett.}}\
  }%
  \textbf{\bibinfo {volume} {102}},\ \bibinfo {pages} {091805} (\bibinfo {year}
  {2009}),\ \Eprint{http://arxiv.org/abs/0811.0053}{arXiv:0811.0053 [hep-ex]}%
  \bibAnnoteFile{NoStop}{Aaltonen:2008ah}%
\bibitem{Fogli:2008ig}%
  \BibitemOpen
  \bibfield{author}{%
  \bibinfo {author} {\bibfnamefont{G.~L.}\ \bibnamefont{Fogli}} \emph{et~al.},\
  }%
  \bibfield{journal}{%
  \Doi{10.1103/PhysRevD.78.033010}{\bibinfo {journal} {Phys. Rev.}}\ }%
  \textbf{\bibinfo {volume} {D78}},\ \bibinfo {pages} {033010} (\bibinfo {year}
  {2008}),\ \Eprint{http://arxiv.org/abs/0805.2517}{arXiv:0805.2517 [hep-ph]}%
  \bibAnnoteFile{NoStop}{Fogli:2008ig}%
\bibitem{Pukhov:2004ca}%
  \BibitemOpen
  \bibfield{author}{%
  \bibinfo {author} {\bibfnamefont{A.}~\bibnamefont{Pukhov}}}%
   (\bibinfo {year} {2004}),\
  \Eprint{http://arxiv.org/abs/hep-ph/0412191}{arXiv:hep-ph/0412191}%
  \bibAnnoteFile{NoStop}{Pukhov:2004ca}%
\bibitem{Semenov:1996es}%
  \BibitemOpen
  \bibfield{author}{%
  \bibinfo {author} {\bibfnamefont{A.~V.}\ \bibnamefont{Semenov}}}%
   (\bibinfo {year} {1996}),\
  \Eprint{http://arxiv.org/abs/hep-ph/9608488}{arXiv:hep-ph/9608488}%
  \bibAnnoteFile{NoStop}{Semenov:1996es}%
\bibitem{Gunion:1989we}%
  \BibitemOpen
  \bibfield{author}{%
  \bibinfo {author} {\bibfnamefont{J.~F.}\ \bibnamefont{Gunion}}, \bibinfo
  {author} {\bibfnamefont{H.~E.}\ \bibnamefont{Haber}}, \bibinfo {author}
  {\bibfnamefont{G.~L.}\ \bibnamefont{Kane}},\ and\ \bibinfo {author}
  {\bibfnamefont{S.}~\bibnamefont{Dawson}},\ }%
  \emph{\bibinfo {title} {{THE HIGGS HUNTER'S GUIDE}}}\ (\bibinfo {publisher}
  {Addison Wesley},\ \bibinfo {year} {1990})%
  \bibAnnoteFile{NoStop}{Gunion:1989we}%
\bibitem{Graudenz:1992pv}%
  \BibitemOpen
  \bibfield{author}{%
  \bibinfo {author} {\bibfnamefont{D.}~\bibnamefont{Graudenz}}, \bibinfo
  {author} {\bibfnamefont{M.}~\bibnamefont{Spira}},\ and\ \bibinfo {author}
  {\bibfnamefont{P.~M.}\ \bibnamefont{Zerwas}},\ }%
  \bibfield{journal}{%
  \Doi{10.1103/PhysRevLett.70.1372}{\bibinfo {journal} {Phys. Rev. Lett.}}\ }%
  \textbf{\bibinfo {volume} {70}},\ \bibinfo {pages} {1372} (\bibinfo {year}
  {1993})%
  \bibAnnoteFile{NoStop}{Graudenz:1992pv}%
\bibitem{Spira:1995rr}%
  \BibitemOpen
  \bibfield{author}{%
  \bibinfo {author} {\bibfnamefont{M.}~\bibnamefont{Spira}}, \bibinfo {author}
  {\bibfnamefont{A.}~\bibnamefont{Djouadi}}, \bibinfo {author}
  {\bibfnamefont{D.}~\bibnamefont{Graudenz}},\ and\ \bibinfo {author}
  {\bibfnamefont{P.~M.}\ \bibnamefont{Zerwas}},\ }%
  \bibfield{journal}{%
  \Doi{10.1016/0550-3213(95)00379-7}{\bibinfo {journal} {Nucl. Phys.}}\ }%
  \textbf{\bibinfo {volume} {B453}},\ \bibinfo {pages} {17} (\bibinfo {year}
  {1995}),\ \Eprint{http://arxiv.org/abs/hep-ph/9504378}{arXiv:hep-ph/9504378}%
  \bibAnnoteFile{NoStop}{Spira:1995rr}%
\bibitem{Barate:2003sz}%
  \BibitemOpen
  \bibfield{author}{%
  \bibinfo {author} {\bibfnamefont{R.}~\bibnamefont{Barate}} \emph{et~al.}
  (\bibinfo {collaboration} {LEP Working Group for Higgs boson searches}),\ }%
  \bibfield{journal}{%
  \Doi{10.1016/S0370-2693(03)00614-2}{\bibinfo {journal} {Phys. Lett.}}\ }%
  \textbf{\bibinfo {volume} {B565}},\ \bibinfo {pages} {61} (\bibinfo {year}
  {2003}),\ \Eprint{http://arxiv.org/abs/hep-ex/0306033}{arXiv:hep-ex/0306033}%
  \bibAnnoteFile{NoStop}{Barate:2003sz}%
\bibitem{Perez:2009mu}%
  \BibitemOpen
  \bibfield{author}{%
  \bibinfo {author} {\bibfnamefont{P.}~\bibnamefont{Fileviez~Perez}}, \bibinfo
  {author} {\bibfnamefont{T.}~\bibnamefont{Han}},\ and\ \bibinfo {author}
  {\bibfnamefont{T.}~\bibnamefont{Li}},\ }%
  \bibfield{journal}{%
  \Doi{10.1103/PhysRevD.80.073015}{\bibinfo {journal} {Phys. Rev.}}\ }%
  \textbf{\bibinfo {volume} {D80}},\ \bibinfo {pages} {073015} (\bibinfo {year}
  {2009}),\ \Eprint{http://arxiv.org/abs/0907.4186}{arXiv:0907.4186 [hep-ph]}%
  \bibAnnoteFile{NoStop}{Perez:2009mu}%
\bibitem{Asner:2010ve}%
  \BibitemOpen
  \bibfield{author}{%
  \bibinfo {author} {\bibfnamefont{D.~M.}\ \bibnamefont{Asner}} \emph{et~al.}}%
   (\bibinfo {year} {2010}),\
  \Eprint{http://arxiv.org/abs/1004.0535}{arXiv:1004.0535 [hep-ph]}%
  \bibAnnoteFile{NoStop}{Asner:2010ve}%
\bibitem{bbmp}%
  \BibitemOpen
  \bibfield{author}{%
  \bibinfo {author} {\bibfnamefont{L.}~\bibnamefont{Basso}}, \bibinfo {author}
  {\bibfnamefont{A.}~\bibnamefont{Belyaev}}, \bibinfo {author}
  {\bibfnamefont{S.}~\bibnamefont{Moretti}},\ and\ \bibinfo {author}
  {\bibfnamefont{G.~M.}\ \bibnamefont{Pruna}}}%
   (\bibinfo {year} {2011}),\ \bibinfo {note} {in progress}%
  \bibAnnoteFile{NoStop}{bbmp}%
\bibitem{CuhadarDonszelmann:2008jp}%
  \BibitemOpen
  \bibfield{author}{%
  \bibinfo {author} {\bibfnamefont{T.}~\bibnamefont{Cuhadar-Donszelmann}},
  \bibinfo {author} {\bibfnamefont{M.}~\bibnamefont{Karagoz}}, \bibinfo
  {author} {\bibfnamefont{V.~E.}\ \bibnamefont{Ozcan}}, \bibinfo {author}
  {\bibfnamefont{S.}~\bibnamefont{Sultansoy}},\ and\ \bibinfo {author}
  {\bibfnamefont{G.}~\bibnamefont{Unel}},\ }%
  \bibfield{journal}{%
  \Doi{10.1088/1126-6708/2008/10/074}{\bibinfo {journal} {JHEP}}\ }%
  \textbf{\bibinfo {volume} {10}},\ \bibinfo {pages} {074} (\bibinfo {year}
  {2008}),\ \Eprint{http://arxiv.org/abs/0806.4003}{arXiv:0806.4003 [hep-ph]}%
  \bibAnnoteFile{NoStop}{CuhadarDonszelmann:2008jp}%
\end{thebibliography}%

\end{document}